\documentclass[a4paper, 11pt, oneside]{article}
\pdfoutput=1
\usepackage{aapreprint}
\makeatletter
\def\@fpheader{\hfill MITP/16--138}
\makeatother

\usepackage{graphicx,wrapfig,float,slashed}
\usepackage{amsmath,amssymb,dsfont,epsfig,graphicx,xcolor}
\usepackage{epstopdf}
\usepackage{ragged2e}
\usepackage[utf8]{inputenc} 
\usepackage{cancel}
\usepackage[table]{xcolor}
\usepackage[toc]{appendix}
\usepackage[normalem]{ulem}
\usepackage[font=small,skip=1pt]{caption}
\allowdisplaybreaks
\usepackage{pifont}
\usepackage{subcaption}
\usepackage{footnote}

\newcommand{\nc}{\newcommand}
\nc{\non}{\nonumber}
\nc{\hc}{\hbox {h.c.}}
\nc{\noi}{\noindent}
\nc{\barx}{\bar{x}}
\nc{\pbarn}{\;\hbox {pb}}
\nc{\fbarn}{\;\hbox {fb}}

\nc{\hsp}{\hspace{0.5cm}}
\nc{\lsp}{\hspace{1cm}}
\nc{\Lsp}{\hspace{2cm}}
\nc{\LLsp}{\lsp\lsp}
\nc{\lra}{\longrightarrow}
\nc{\p}{\prime}
\nc{\sgn}{\text{sgn}}
\nc{\tr}{\text{Tr}}
\nc{\ph}{\varphi}
\nc{\op}{{\cal O}}
\nc{\cL}{{\cal L}}
\nc{\cU}{{\mathcal U}}
\nc{\cD}{{\mathcal D}}

\nc{\beq}{\begin{equation}}  \nc{\eeq}{\end{equation}}
\nc{\bea}{\begin{eqnarray}}  \nc{\eea}{\end{eqnarray}}
\nc{\baa}{\begin{array}}     \nc{\eaa}{\end{array}}
\nc{\bit}{\begin{itemize}}   \nc{\eit}{\end{itemize}}
\nc{\ben}{\begin{enumerate}} \nc{\een}{\end{enumerate}}
\nc{\bce}{\begin{center}}    \nc{\ece}{\end{center}}
\nc{\bpm}{\begin{pmatrix}}   \nc{\epm}{\end{pmatrix}}
\nc{\bvt}{\begin{verbatim}}  \nc{\evt}{\end{verbatim}}

\def\lsim{\mathrel{\raise.3ex\hbox{$<$\kern-.75em\lower1ex\hbox{$\sim$}}}}
\def\gsim{\mathrel{\raise.3ex\hbox{$>$\kern-.75em\lower1ex\hbox{$\sim$}}}}

\def\udots{\mathinner{\mkern1mu\raise1pt\vbox{\kern7pt\hbox{.}}\mkern2mu\raise4pt\hbox{.}\mkern2mu\raise7pt\hbox{.}\mkern1mu}}

\definecolor{agray}{rgb}{0.95, 0.95, 0.99}

\def\eg{{\it e.g.}}

\def\cN{\mathcal{N}}

\def\rcite#1{Ref.~\cite{#1}}

\title{\huge Clockwork Goldstone Bosons}
\author[1,2]{Aqeel Ahmed}
\emailAdd{aqeel.ahmed@fuw.edu.pl}
\affiliation[1]{Faculty of Physics,
University of Warsaw,
Pasteura 5, 02-093 Warsaw, Poland}
\affiliation[2]{Mainz Institute for Theoretical Physics, Johannes Gutenberg University, 55099 Mainz, Germany}
\author[3]{and Barry M. Dillon}
\emailAdd{barry.dillon@plymouth.ac.uk}
\affiliation[3]{Centre for Mathematical Sciences, Plymouth University, PL4-8AA Plymouth, UK}
\date{\today}

\abstract{
The clockwork mechanism has recently been proposed as a natural way to generate hierarchies among parameters in quantum field theories. 
The mechanism is characterized by a very specific pattern of spontaneous and explicit symmetry breaking, and the presence of new light states referred to as `gears'.
In this paper we begin by investigating the self-interactions of these gears in a scalar clockwork model and find a parity-like selection rule at all orders in the fields.
We then proceed to investigate how the clockwork mechanism can be realized in 5D linear dilaton models from the spontaneous symmetry breaking of a complex bulk scalar field.
We also discuss how the clockwork mechanism is manifest in the scalar components of 5D gauge theories in the linear dilaton model, and build their 4D deconstructed analogue.
Finally we discuss attempts at building both 4D and 5D realizations of a non-abelian scalar clockwork mechanism, where in the latter we consider scenarios in which the Goldstone bosons arise from 5D scalar and 5D gauge fields.}

\begin{document}

\maketitle
\flushbottom


\section{Introduction} \label{introduction}

Understanding phenomena at different energy scales and how the hierarchies in these scales originate has been one of the major driving forces in Beyond the Standard Model (BSM) physics. 
The recently proposed clockwork mechanism is one way in which hierarchies may be generated between couplings and scales in certain models.
They were first proposed in the context of axion (relaxion) models \cite{Choi:2015fiu,Kaplan:2015fuy}, and then later generalized to models involving particles of different spins \cite{Giudice:2016yja}, most notably using a clockwork mechanism for gravitons as a solution to the hierarchy problem. There have since been various applications of the clockwork mechanism, in particular to natural inflation \cite{Kehagias:2016kzt} and other axion models \cite{Farina:2016tgd}. 

The simplest realization of a clockwork mechanism involves a set of scalar fields arranged in a lattice configuration with nearest-neighbor interactions only.
A specific pattern of spontaneous and explicit breaking of the global symmetries leads to one exact global clockwork symmetry which ensures a massless Goldstone boson whose properties are completely fixed by this clockwork symmetry, in particular its interactions to external fields can be naturally suppressed by exponentially small couplings.
It was observed in \cite{Giudice:2016yja} that the continuum limit of these discrete lattice models can be approximated quite well by 5D fields in a linear dilaton (LD) background geometry \cite{Antoniadis:2011qw}, see also \cite{Baryakhtar:2012wj,Cox:2012ee,Ahmed:2013lea}.
This 5D model is compactified on an interval between two 3-branes, thus 5D fields in the bulk of the extra dimension can be expanded in a series of 4D mass eigenstates, where the clockwork Goldstone boson is identified with a massless component of a bulk scalar field.
The clockwork-suppressed interactions arise between this Goldstone field, which lives in the bulk, and fields localized to one the 3-branes.

A crucial ingredient in generating the clockwork mechanism is a very specific form of mass mixing between the neighboring sites.
In Sec.~\ref{quartics} we study the higher order interactions among the scalar fields which arise naturally and automatically from the clockwork model.
Interestingly we find an analytic expression for these interactions in the mass eigenbasis, and a parity-like selection rule determining the structure of these interactions to all orders in the fields.

In Sec.~\ref{abeliansec} we begin by studying a realization of the clockwork mechanism in the LD model which consists of a complex scalar field in the bulk of the extra dimension with no brane potentials.
Allowing this field to acquire a VEV which spontaneously breaks the global $U(1)$ symmetry in the bulk, we study the properties of the radial and longitudinal fluctuations under a Kaluza-Klein (KK) decomposition.
We find that only the longitudinal fluctuation is allowed to have a zero mode, this corresponds to the true Goldstone boson of the spontaneous symmetry breaking.
And interestingly, we find that there is a natural mass gap between the radial and longitudinal KK modes due to the bulk mass of the radial component.
This feature is not present in other warped models such as Randall-Sundrum models \cite{Randall:1999ee}, and is very similar to the separation of scales between the radial and longitudinal modes in the discrete case.
The couplings of the Goldstone boson in the 5D model are clockwork-suppressed with respect to those of the massive KK modes when the couplings are to fields localized on a 3-brane.
We then study a simple model with an abelian gauge field in the 5D bulk and boundary conditions chosen such that the $A_5$ zero mode remains in the spectrum while the gauge zero mode is projected out.
This $A_5$ zero mode can be identified with the Goldstone boson of the spontaneous breaking of the bulk gauge symmetry on the branes. 
Similar mechanisms have been used in other 5D models to study the Higgs field in the Standard Model electroweak sector, where the mechanism goes by the name gauge-Higgs unification \cite{Hosotani:1983xw,Contino:2003ve,Agashe:2003zs,Agashe:2004rs,Hosotani:2005nz}.
Interestingly, the properties of this $A_5$ are strikingly similar to those of the bulk complex scalar.
In fact the so-called `physical localizations' of these fields in the 5D bulk are related by flipping the sign of the curvature constant.
We find that this $A_5$ mode is another realization of a scalar clockwork model, and we proceed to build a deconstructed 4D version of this model consisting of $U(1)$ gauge fields and scalar fields.
Because the clockwork parameter arises from an abelian gauge coupling this model naturally allows for non-integer values of the coupling $q$.

In the last section of the paper we discuss attempts at building a non-abelian clockwork model for scalars~\footnote{Since the first version of this manuscript appeared on arXiv, further discussions of non-abelian clockwork models have appeared in the literature \cite{Craig:2017cda}, see also \cite{Ben-Dayan:2017rvr}.}.
We begin by discussing the discrete case, and conclude that one cannot have a multiplet of exact Goldstone bosons while also generating a clockwork mechanism due to the presence of terms that explicitly break the global symmetry.  However it is known that these terms go to zero in the continuum limit, thus we proceed to study a 5D realization of a non-abelian clockwork model.
The model is a simple extension of the abelian case, where we have a multiplet of complex scalar fields in the bulk of the LD model with an $SU(2)_L\times SU(2)_R$ global symmetry.
The results here follow in the exact same way as in the abelian case.
Finally we discuss non-abelian bulk gauge fields in the LD model, a generalization of the abelian gauge model.
The model consists of a bulk gauge symmetry broken by boundary conditions such that some combination of massless $A_5$ scalars (corresponding again to Goldstone bosons) and gauge fields survive.
We find that the pattern of couplings between the different fields depends on the sign of the curvature term in 5D, with one sign providing a suppression of the gauge coupling and the other providing a suppression of couplings to the Goldstone bosons.
Moreover, we supplement the main results with two appendices, where Appendix~\ref{appendixA} contains a discussion of higher dimensional operators in the bulk of the 5D theory. Whereas in Appendix~\ref{gauge_LDM} we demonstrate the bulk gauge symmetry breaking and KK-decomposition of the bulk gauge theories in the Jordan and Einstein frames of LD model.

\subsection{Abelian clockwork} \label{clockwork}

In this subsection we review the underlying mechanism for the clockwork theory with a $U(1)^{N\!+\!1}$ global symmetry. Following \cite{Choi:2015fiu,Kaplan:2015fuy,Giudice:2016yja} we take the Lagrangian for  $N\!+\!1$ complex scalars
\beq \label{axionpotential}
\mathcal{L}=\sum_{j=0}^{N}\Big(\partial_{\mu}\phi_j^\dag\partial^{\mu}\phi_j-\mu^2\phi_j^{\dagger}\phi_j-\frac{\lambda}{4}\vert\phi_j^{\dagger}\phi_j\vert^2\Big)+\epsilon^2\tilde m^{3-q}\sum_{j=0}^{N-1}\Big(\phi_j^{\dagger}\phi_{j+1}^q + \hc\Big),
\eeq
where $q\!>\!1$ (and $q\!\in\! \mathbb{Z}$) and  $\epsilon$ are the dimensionless parameters which control explicit breaking of the global symmetries, and $\tilde m$ has dimension of mass.
Throughout this work we refer to each of the $j$ values as a {\it site}. 
The first part of the action contains interactions located at a particular site, whereas the second part provides links, or couplings, between the sites.
The first part of the Lagrangian is invariant under a global $U(1)^{N\!+\!1}$ symmetry defined by
\beq
\phi_j\rightarrow e^{i\theta_j}\phi_j ~,
\eeq
for each field.
However, under this transformation the second part of the potential transforms to
\beq
\epsilon^2\tilde m^{3-q}\sum_{j=0}^{N-1}\Big(\phi_j^{\dagger}\phi_{j+1}^qe^{-i(\theta_j-q\theta_{j+1})} + \hc\Big),
\eeq
thus this second part is only invariant under one global symmetry $U(1)^{\prime}$ in which 
\beq \label{u1charges}
\theta_0=q\theta_1=q^2\theta_2=\cdots=q^N\theta_{N}.
\eeq
Assuming $\mu^2\!<\!0$ the scalar fields $\phi_j$ will get vacuum expectation values
\beq
v_\phi=\sqrt\frac{-2\mu^2}{\lambda}~.
\eeq
In the limit $\epsilon^2\rightarrow0$ we see that these VEV's spontaneously break the $U(1)^{N\!+\!1}$ symmetry.
Each $\phi_j$ has two degrees of freedom, one of which gets a mass of $m_\phi^2\!=\!\lambda v_\phi^2$, while the second is the Goldstone boson ($\pi_j$) which remains massless.
The mass scale of the theory here is set by $v_\phi$, however we also have two dimensionless couplings $\epsilon$ and $\lambda$.
If we assume $\epsilon^2\ll \lambda$ then we essentially have two mass scales, one set by $\lambda v_\phi^2$ and one by $\epsilon^2\tilde m^2 \!\simeq\!\epsilon^2 v_\phi^2$.

At scales $\sim\! \epsilon^2v_\phi^2\!\ll\!\lambda v_\phi^2$, we can use a non-linear sigma description of the Goldstone fields,
\beq \label{ujdef}
\phi_j\rightarrow  f_{\pi}U_j\equiv  f_{\pi}e^{i\pi_j/\sqrt{2}f_{\pi}}~,
\eeq
where $f_{\pi}\!\sim\! v_\phi$ is the decay constant of the Goldstone bosons.
We write the Lagrangian for these states as
\beq
\mathcal{L}=f_{\pi}^2\sum_{j=0}^N\partial_{\mu}U_j^{\dagger}\partial^{\mu}U_j+\epsilon^2 f_{\pi}^4\sum_{j=0}^{N-1}\left( U_j^{\dagger}U_{j+1}^q + \text{h.c.}  \right),		\label{cw_eft}
\eeq
where for simplicity we take $\tilde m\!=\!f_\pi$. 
It is important to comment here that if we only consider the above effective field theory (EFT) then $q\in\mathbb{R}$, i.e. it is not required to be an integer, which is a crucial aspect if one wants to consider a continuum description of this EFT model, see also \cite{Giudice:2016yja}.
In the limit $\epsilon^2\rightarrow 0$ these Goldstone fields have no potential at any order in perturbation theory.
However non-zero $\epsilon$ introduces a potential at tree-level through the mixing terms.
Expanding
\beq
U_j\simeq 1 +\frac{i}{\sqrt{2}f_{\pi}}\pi_j~,
\eeq
and plugging this into the above Lagrangian~(\ref{cw_eft}) we see that terms odd in $\pi_j$ cancel with the hermitian conjugate and the lowest order potential is the quadratic piece,
\beq
\mathcal{L}_\pi=\frac{1}{2}\sum_{j=0}^N\partial_{\mu}\pi_j^{\dagger}\partial^{\mu}\pi_j-\frac{1}{2}\epsilon^2 f_{\pi}^2\sum_{j=0}^{N-1}(\pi_{j}-q\pi_{j+1})^2+\mathcal{O}(\pi^4).
\eeq
We expect this explicit breaking term to break the non-linearly realized $U(1)^{N\!+\!1}$ symmetry to the $U(1)^{\prime}$ described in Eq.~(\ref{u1charges}), giving $N$ of the Goldstone bosons mass and leaving one massless.

The potential terms at the quadratic order can be written as
\beq
V^{(2)}(\pi)=\frac{1}{2}\epsilon^2 f_{\pi}^2 \pi_i M^2_{ij} \pi_j~,
\eeq
where the sums over the $i,j$ indices in the above expression are assumed to be over $0,\ldots,N$. We will always assume this summation convention when we have repeated indices, if the sum is not over this interval we will write the summation explicitly. The mass matrix above has the following form
\beq \label{clockworkMassM}
M_{ij}^2=\bpm
1 & -q & 0 & \cdots &  0& 0 \cr 
-q & 1+q^2 & -q & \cdots &0  & 0 \cr
0 & -q & 1+q^2 & \cdots & 0& 0 \cr
\vdots & \vdots & \vdots & \ddots & \vdots&\vdots \cr
 0& 0&0 & \cdots& 1+q^2 & -q \cr
 0 & 0 & 0 &\cdots & -q & q^2
\epm.
\eeq
This is diagonalised by an orthogonal transformation given by
\beq \label{ClockworkRot}
\mathcal{O}_{j0}=\frac{\mathcal{N}_0}{q^j},\lsp \mathcal{O}_{jk}=\mathcal{N}_k\left(q\sin\frac{jk\pi}{N\!+\!1}-\sin\frac{(j+1)k\pi}{N\!+\!1}\right),
\eeq
where $j=0,\ldots,N$ and $k=1,\ldots,N$,
and 
\beq\label{cweq3}
\mathcal{N}_0=\sqrt{\frac{q^2-1}{q^2-q^{-2N}}}~,\lsp \mathcal{N}_k=\sqrt{\frac{2}{(N\!+\!1)\lambda_k}}~,\lsp \lambda_k=q^2+1-2q\cos\Big(\frac{k\pi}{N\!+\!1}\Big).
\eeq
The eigenvalues, i.e. the masses of the pseudo-Goldstone modes, are positive-definite and given by
\beq \label{mass_sq}
m_k^2=\lambda_k\epsilon^2 f_{\pi}^2, \lsp\text{for}\hsp k=1,\ldots,N
\eeq
whereas the $k\!=\!0$ mode remains massless and is the true Goldstone of the spontaneously broken global $U(1)^\p$ symmetry.
The form of $\mathcal{O}_{j0}$ can immediately be understood by inspecting the mixing term after diagonalization
\beq \label{pi0int}
\epsilon^2f_\pi^4\left(e^{-\frac{i}{\sqrt{2}f_{\pi}}(\pi_j-q\pi_{j+1})} +\text{h.c.}\right) \rightarrow  \epsilon^2f_\pi^4\left(e^{-\frac{i}{\sqrt{2}f_{\pi}}(\mathcal{O}_{jk}-q\mathcal{O}_{(j+1)k})\pi_k^\p}+\text{h.c.}\right).
\eeq
For $\pi_0^\p$ to be a true Goldstone it must not have any direct couplings to the heavier modes.  
Thus this interaction term in Eq.~(\ref{pi0int}) must be independent of $\pi_0^\p$ after diagonalization, which requires $\mathcal{O}_{j0}=q\mathcal{O}_{(j+1)0}$, as in Eq.~(\ref{ClockworkRot}).
The diagonalized Lagrangian is now
\beq
\mathcal{L}_{\pi^\prime}=\frac{1}{2}\sum_{j=0}^N\left(\partial_{\mu}\pi_j^{\prime\dagger}\partial^{\mu}\pi^{\prime}_j-m_j^2\pi_j^{\p2}\right) + \mathcal{O}(\pi^{\prime4})~,
\eeq
where
\beq \label{expans}
\pi_j=\mathcal{O}_{jk}\pi^{\prime}_k.
\eeq
We refer to the $\pi^\p$ fields as being in the clockwork basis, and the $\pi$ fields as being in the interaction basis.
In the example of this section the clockwork basis and the mass eigenbasis are the same, but if there are other interactions which mix the scalar fields at each site then this may not necessarily be true.
In the large $N$ limit the mass range of these massive modes so-called `gears' is given by
\beq
m_N^2-m_1^2=4q\epsilon^2 f_{\pi}^2,
\eeq
thus the spacing between modes must become very small.  
With $q\!>\!1$ we find $\mathcal{O}_{j0}\!\sim\! q^{-j}$ which then grows smaller for larger values of $j$.
This implies that the size of the $j^{\text{th}}$ scalar in $\pi^{\prime}_0$ gets smaller with larger $j$ and in the limit $q\!\rightarrow\! \infty$ we have $\pi^{\prime}_0\!=\!\pi_0$.
When $q\!<\!1$ we find $\mathcal{O}_{j0}\sim q^{N\!-\!j}$ which implies the opposite.
That is the size of the $j^{\text{th}}$ scalar in the diagonalised Goldstone gets larger as with larger $j$. 

Now we take $q\!>\!1$ and couple a fermion field $\psi$ to the $N^{\text{th}}$ site with the $\pi_N$ field, this interaction gets transferred to the massless Goldstone after the diagonalisation of the terms $\sim \epsilon^2$ and we find,
\begin{align}
\mathcal{L}_{\text{int}}= -Y\frac{\pi_N}{f_\pi}\bar{\psi}\psi&= -\mathcal{O}_{N0}Y\frac{\pi_0^{\prime}}{f_\pi}\bar{\psi}\psi-\sum_{k=1}^N\mathcal{O}_{Nk}Y\frac{\pi_k^{\prime}}{f_\pi}\bar{\psi}\psi~, \nonumber \\
&\simeq- Y\frac{\pi_0^{\prime}}{f_\pi q^{N}}\bar{\psi}\psi-\sum_{k=1}^NY\frac{\pi_k^{\prime}}{f_\pi}\bar{\psi}\psi\, ,
\end{align}
with the normalisation constants ($\cN_0$ and $\cN_{k}$) being $\mathcal{O}(1)$.
Thus the interaction between the fermion field and the exact Goldstone boson will be suppressed by $q^{N}$ as compared to the {\it gears}' interactions.
These are the {\it gears} of the clockwork mechanism through them that the hierarchical couplings are being generated.
Coupling a gauge or fermion field to all scalars with the same couplings is invariant under the clockwork rotation, thus results in no suppression of the couplings at all.

\section{Clockwork parity:~{\bf accidental~symmetries~in~clockwork~interactions}} \label{quartics}

In this section we consider the higher order clockwork interactions in the effective action. 
We consider the EFT description of the clockwork model with the potential of the following form, 
\beq  
V(\pi)=-\epsilon^2 f_{\pi}^4\sum_{j=0}^{N-1}\left( U_j^{\dagger}U_{j+1}^q + \text{h.c.}  \right),
\eeq
where $U_j$ is defined in Eq.~(\ref{ujdef}).
The above potential can be written as,
\begin{align} \label{cospot}
V(\pi)&=-2\epsilon^2 f_{\pi}^4\sum_{j=0}^{N-1}\cos\left(\frac{\pi_j-q\pi_{j+1}}{\sqrt{2}f_{\pi}}\right),	\notag\\
&=V_0+\sum_{j=0}^{N-1}\left[\frac{\epsilon^2 f_{\pi}^2}{2}\left(\pi_j-q\pi_{j+1}\right)^2-\frac{\epsilon^2}{2 (4!)}\left(\pi_j-q\pi_{j+1}\right)^4\right]+\op(\pi^6),
\end{align}
where $V_0\!\equiv\!-2\epsilon^2 f_{\pi}^4$, is a constant vacuum energy density that can be absorbed in the definitions of the fields such that we can set $V_0=0$, without loss of generality. In order to get the physical interactions we rotate the $\pi_j$ fields to the clockwork basis as in Eq.~(\ref{expans}).
The above potential then has the following form up to quartic level in fields, 
\begin{align} \label{piaction}
V(\pi)&= \sum_{j=0}^{N-1}\left[\frac{\epsilon^2f_{\pi}^2}{2}R_{jk}R_{jl}\pi^\p_k\pi^\p_l-\frac{\epsilon^2}{2(4!)}R_{jk}R_{jl}R_{jm}R_{jn}\pi^\p_k\pi^\p_l\pi^\p_m\pi^\p_n\right], 
\end{align}
where the sums over the $k,l,m,n$ indices are assumed to be over $0,\ldots,N$.
The rotation matrices in Eq.~\eqref{piaction} are defined by
\beq \label{Rjk}
R_{jk}\equiv\mathcal{O}_{jk}-q\mathcal{O}_{j+1,k}
\eeq
such that $R_{j0}=0$ and have the property
\begin{align}\label{diagcond}
\sum_{j=0}^{N-1}R_{jk}R_{jl}&= \delta_{kl} \lambda_k, \hsp\text{for}\hsp l=1,\cdots,N
\end{align}
where $\lambda_k$ is defined in \eqref{cweq3}.
We find that the terms quartic in $R_{jk}$ are generally factors of the $\lambda_k$ terms.
We will conveniently denote them as
\beq
\sum_{j=0}^{N-1}R_{jk}R_{jl}R_{jm}R_{jn}\equiv {\cal C}_{klmn},  \hsp\text{for}\hsp k,l,m,n=0,1,\cdots,N,
\eeq
which is exactly zero when any of the indices equals zero, indicating that there are no quartic interactions for the massless mode.
These terms are pre-factors to interactions which mix the $k^{\text{th}}$, $l^{\text{th}}$, $m^{\text{th}}$, and $n^{\text{th}}$ gears.
We find the following compact form of ${\cal C}_{klmn}$:
\beq
{\cal C}_{klmn}=\frac{4\sqrt{\lambda_k\lambda_l\lambda_m\lambda_n}}{(N+1)^2}\sum_{s=0}^{N}\sin\left(\frac{s k\pi}{N+1}\right)\sin\left(\frac{s l\pi}{N+1}\right)\sin\left(\frac{s m\pi}{N+1}\right)\sin\left(\frac{s n\pi}{N+1}\right).\label{Cklmn}
\eeq
One can in fact find analytic expressions for the coefficients of $d$-number of self-interacting pseudo-Goldstone fields $\pi^\p_j$ in terms of $R_{jk}$, summed over $j$, as,
\begin{align} 
\sum_{j=0}^{N-1}R_{jk_1}R_{jk_2}\ldots R_{jk_{d}}&=(-1)^d\left(\frac{2}{N+1}\right)^{d/2}\sqrt{\lambda_{k_1}\lambda_{k_2}\ldots\lambda_{k_{d}}} \nonumber \\
&\quad\times\sum_{s=0}^{N}\sin\left(\frac{s k_1\pi}{N+1}\right)\sin\left(\frac{s k_2\pi}{N+1}\right)\ldots\sin\left(\frac{s k_{d}\pi}{N+1}\right), \nonumber \\
&=(-1)^d\left(\frac{2}{N+1}\right)^{d/2}\sum_{s=0}^{N}\prod_{t=1}^d\sqrt{\lambda_{k_{t}}}\sin\left(\frac{s k_t\pi}{N+1}\right).  \label{parity}
\end{align}
Note that setting $d\!=\!2$ in the above expression one finds $\lambda_{k_1}\delta_{k_1k_2}$, as required to diagonalize the mass matrix.
Because of the structure of the clockwork interaction terms, we only have interactions with even $d$, however the above expression works equally well for odd $d$.
The interesting feature here is that the terms inside the sum are now independent of $q$, all the $q$ dependence has been factored out into the $\lambda_k$ terms.

These interactions contain a very interesting structure, for example the quartic term can be written as
\begin{align}\label{Cklmn2}
{\cal C}_{klmn}=\frac{\sqrt{\lambda_k\lambda_l\lambda_m\lambda_n}}{2(N+1)}&\Big[\hat\delta_{k+l+m+n}+\hat\delta_{k+l-m-n}-\hat\delta_{k+l+m-n} -\hat\delta_{k+l-m+n}		\\ \nonumber 
&-\hat\delta_{k-l+m+n}-\hat\delta_{k-l-m-n}+\hat\delta_{k-l+m-n}+\hat\delta_{k-l-m+n}\Big]
\end{align}
where $\hat\delta_{\mathbf{z}}$ is defined as,
\beq
\hat\delta_{\mathbf{z}}\equiv \left\{\begin{array}{l}
1 	\lsp {\rm for }\hsp \mathbf{z}=0\text{~mod~}2(N\!+\!1)\\
0	\lsp {\rm for }\hsp \mathbf{z}\neq0\text{~mod~}2(N\!+\!1)
\end{array}
\right.\, .
\eeq
This is very reminiscent of the KK parity selection rules found in extra dimensional models.
In the large-$N$ limit, or simply when $k,l,m,n<(N+1)/2$, we have exact KK parity.
The rule actually extends to all of the higher order interactions of the gears, thus the selection rule can be stated for Eq.~\eqref{parity} such that
\beq \label{parity2}
k_1\pm k_2\pm k_3\pm \ldots\pm k_d=0~\text{mod}~2(N+1).
\eeq
It is also worth noting that these interactions are suppressed by increasing powers of $N\!+\!1$, depending on the number of fields in the interaction.
Thus with a large number of sites, even the interactions involving only light gears are suppressed.


\section{Abelian clockwork models from extra dimension}  
\label{abeliansec}
In this section we will present a study of the spontaneous breaking of global $U(1)$ symmetries in the bulk of a 5D linear dilaton model.
We will see that natural choices of 5D parameters can result in an EFT which has all the clockwork properties. In particular the effective 4D couplings turn out to be exponentially suppressed, and the mass spectrum of the heavy modes matches that of the gears in the discrete clockwork model.

\subsection{A clockwork scalar from a 5D complex scalar}

The linear dilaton model is a compactified extra dimensional model, where the extra dimensional space is bounded between two 3-branes at positions $y=0$ and $y=L$ along the extra dimension \cite{Antoniadis:2011qw} (see also \cite{Giudice:2016yja,Baryakhtar:2012wj,Cox:2012ee}).
In the so-called Jordan frame of the linear dilaton model we can write the action for a complex scalar field $\phi(x,y)$ as
\begin{align}
S=\!\int \!d^4x\int_0^L\!dy~e^{2ky}\left[|\partial_M\phi|^2-\mu^2|\phi|^2-\frac{\lambda}{4}|\phi|^4\right],	\label{salar5DAction}
\end{align}
where we have included terms leading to the spontaneous breaking of the global $U(1)$ symmetry in the bulk. The indices above are contracted with the 5D Minkowski metric and the mass dimensions of the fields and couplings are: $[\phi]=\tfrac32$, $[\mu]=[k]=1$, and $[\lambda]=-1$.
We have not included brane interaction terms here, the rationale we use to justify this is that in the clockwork model the potential at each site is equivalent, and in the continuum limit the fields at each site become the field values along the extra dimensional coordinate.
It is straightforward to extend the model to include brane mass terms, however this is beyond the scope of what we want to demonstrate here.

For $\mu^2\!<\!0$ the field $\phi$ acquires a VEV, say $\langle\phi(x,y)\rangle\!=\!v(y)$. The solution for $v(y)$ can be obtained by minimizing the 5D action,
\begin{align}
-\partial_5\left(e^{2ky}~\partial_5v(y)\right)+\frac{\partial V(\phi)}{\partial \phi}\bigg\vert_{\phi=v}\!=0,	\lsp{\rm where} \hsp V(\phi)\!=\!\mu^2|\phi|^2+\frac{\lambda}{4}|\phi|^4,	\label{eom_v}
\end{align}
along the boundary condition \begin{align}
e^{2ky}v(y)\partial_5v(y)\Big\vert_{0,L}&=0. \label{v_bc}
\end{align}
Moreover, we introduce the normalized VEV as, 
\beq
\int_0^L\!dy v(y)^2=f_\pi^2,	\label{v_norm}
\eeq
where the mass dimensions of $[f_\pi]\!=\!1$. 
In order to have non-zero VEV in the absence of brane-localized term we consider the Neumann boundary condition, which implies the VEV is simply a constant and from Eqs.~\eqref{eom_v}--\eqref{v_norm} it follows,
\beq
f_{\pi}=\sqrt{\frac{-2\mu^2L}{\lambda}}~.
\eeq 
Writing the $\phi(x,y)$ field in non-linear sigma form, i.e. 
\beq
\phi(x,y)=\tfrac{1}{\sqrt{2L}}\big[f_{\pi}+\varphi(x,y)\big]U(x,y),	\lsp {\rm with} \lsp U\equiv e^{i\pi(x,y)/f_{\pi}},		\label{phi_decomposition}
\eeq 
where the radial mode $\varphi(x,y)$ and the longitudinal mode $\pi(x,y)$ have mass dimensions 1. Now we can rewrite the 5D action \eqref{salar5DAction} up to quadratic level in fields as 
\beq
S=\int \!d^4x\int_0^L\!dy\frac{e^{2ky}}{2L}\Big[(\partial_M\varphi)^2-\mu^2\varphi^2+(\partial_M\pi)^2\Big].
\eeq
Looking at the radial component $\varphi(x,y)$ first, we perform a KK decomposition
\beq
\varphi(x,y)=e^{-ky}\sum_{n=0}^{\infty}\varphi_n(x)f^{\varphi}_n(y)
\eeq
such that we describe the field in terms of a set of 4D KK-modes and their 5D wave functions. Note that the extra $e^{-ky}$ factor in the KK-decomposition is to make the $f^{\varphi}_n(y)$ ``physically localized'' wave-functions of the KK-modes along the extra dimension.  Normalization of the 4D kinetic term implies 
\beq
\frac{1}{L}\int_0^L\! dyf^{\varphi}_m(y)f^{\varphi}_n(y)=\delta_{mn},
\eeq 
and the condition that the 4D fields be mass eigenstates fixes the 5D profiles to be
\beq
f^{\varphi}_n(y)=\frac{1}{N_n}\Big[\cos\left(\beta_ny\right)+C_n\sin\left(\beta_ny\right)\Big]
\eeq 
where $N_n$ is fixed by the normalization condition and $C_n$ and $\beta_n$ are fixed by the boundary conditions on both branes, i.e. $f^{\varphi}_ne^{ky}\partial_5e^{-ky}g_n\!=\!0|_{0,L}$.
$C_n$ is simply an integration constant whereas $\beta_n\!=\!\sqrt{m_{\varphi,n}^2-(k^2+\mu^2)}$, hence fixing this fixes the mass spectrum of the KK modes.
We choose the same boundary conditions for the KK modes as we choose for the VEV, thus we find
\beq  \label{5Dprofile}
f^{\varphi}_n(y)=\frac{\sqrt2 \,n\pi}{m_{\varphi,n}L}\Big[\cos\left(\frac{n\pi y}{L}\right)+\frac{kL}{n\pi}\sin\left(\frac{n\pi y}{L}\right)\Big], \hsp {\rm with}\hsp m_{\varphi,n}^2=k^2+\mu^2+\left(\frac{n\pi}{L}\right)^2.
\eeq
An interesting feature here is that the mass gap not only depends on the curvature $k$, as one would expect from linear dilaton models, but also on the bulk mass $\mu$.
The $n$-dependent term here is also characteristic of the linear dilaton models, as it allows one to control the mass splittings of the KK modes by varying the size of the extra dimension.
Note that the radial mode $\varphi(x,y)$ does not have a normalizable massless zero-mode solution.
The model gives a bulk mass to the $\varphi(x,y)$ field, and no brane mass terms, therefore it is impossible to find a normalizable massless zero-mode solution that is consistent with the boundary conditions.

Switching our attention now to the longitudinal components $\pi(x,y)$, we perform a KK decomposition on $\pi(x,y)$ as
\beq
\pi(x,y)=e^{-ky}\sum_{n=0}^{\infty}f^\pi_n(y)\pi_n(x),		\label{piKKmodes}
\eeq
where, as mentioned above, the explicit rescaling factor $e^{-ky}$ allows us to interpret the KK wave-functions $f^\pi_n(y)$ as the ``physically localized'' wave-functions along the extra-dimension. 
At the quadratic level the 5D action for the longitudinal field is the same as that for the radial component except that it has no bulk mass term.
We can follow the same steps as we did before but with one crucial difference, the Neumann boundary conditions allow a massless zero-mode solution for the longitudinal field with the 5D profile
\beq
f^\pi_0(y)= N_0~e^{k(y-L)}, \lsp{\rm where}\hsp N_0\equiv \sqrt{\tfrac{2kL}{1-e^{-2kL}}}~.	\label{f0piN0}
\eeq
Along with the massless zero-mode we have a massive tower of KK modes with
\begin{align}
f^\pi_n(y)=\frac{\sqrt2 \, n\pi}{m_{\pi,n}L}\Big[\cos\left(\frac{n\pi y}{L}\right)+\frac{kL}{n\pi}\sin\left(\frac{n\pi y}{L}\right)\Big],	\lsp m_{\pi,n}^2=k^2+\left(\frac{n\pi}{L}\right)^2.
\end{align}
We see that the mass splittings of the longitudinal KK modes is the same as for the radial KK modes, however the mass gap goes as $k$ rather than $\sqrt{k^2+\mu^2}$.
This means that if the bulk mass term is much larger than the curvature, i.e. $\mu\!\gg\! k$, the radial modes can be effectively decoupled from the zero-mode longitudinal mode and the lowest laying longitudinal KK modes.
This is not possible in other warped models like \eg\ Randall-Sundrum model.
In the discrete model we also made a similar assumption in order to neglect effects of the radial modes. Hence, the EFT picture that emerges in the low energy (for $k$ and $\tfrac{n\pi}{L}\ll f_\pi\!\sim\!\mu$) is that we have a zero-mode scalar $\pi_0$ localized toward one end of the extra dimension $(y\!=\!L)$ and a tower of KK-modes $\pi_n$ with flat localization up to a periodic
modulation along the extra-dimension.
This, as we show below, has the same characteristics as the clockwork scalar and the massive gears, respectively.

\subsubsection*{Couplings of the bulk scalar fields} \label{BulkCouplings}

In order to see the continuum clockwork mechanism in action in our setup, we study the coupling of the Goldstone field $\pi(x,y)$ to a fermion bi-linear, where the fermion operator lives on a brane at some point $\hat y$ in the extra dimension.
The interaction on the brane can be written as
\begin{align}
S_{\rm int}&=\int \!d^4x\int_0^L \!dy~\delta(y-\hat y)\Big(i\bar{\psi}\slashed{\partial}\psi-Y\frac{\pi}{f_{\pi}}\bar{\psi}\psi+\text{h.c.}\Big),	\notag\\
&=\int \!d^4x \sum_{n=0}^\infty\Big(i\bar{\psi}\slashed{\partial}\psi-Y\frac{e^{-k\hat y}\pi_n(x) f^\pi_n(\hat y)}{f_{\pi}}\bar{\psi}\psi+\text{h.c.}\Big),
\end{align}
where in the last step we used the KK decomposition and performed the $y$ integral. 
Note that the coupling between the fermion operator and the Goldstone zero-mode and non-zero KK modes are given by
\beq
\mathcal{L}_{\rm int}=-\frac{N_0~Y}{f_{\pi}e^{kL}}\pi_0(x)\bar{\psi}\psi -\frac{Yc_n}{f_\pi e^{k\hat y}}\pi_n(x)\bar{\psi}\psi,
\eeq
where $N_0$ is defined in \eqref{f0piN0} and $c_n$ is an $\mathcal{O}(1)$ coupling dependent on the oscillatory modulations.
From the above interaction Lagrangian, we observe two facts for the zero-mode couplings, 
\bit
\item[(\emph {i})] the zero-mode interaction is $\hat y$-independent, and 
\item[(\emph{ii})] the effective interaction scale is $e^{kL}$ times greater than the fundamental scale $f_\pi$. 
\eit
However the gear interactions depend on the interaction-brane location $\hat y$.
In the case where the location of interaction-brane is at $\hat y\!=\!0$ the interaction of the zero-mode, as compared to the gears (which is $\sim\! f_\pi$), is suppressed by $e^{kL}$. 
On the other hand if we have the interaction brane located at $\hat y\!=\!L$, there is no relative exponential suppression in the couplings since both the zero-mode and gear interactions scale as $f_{\pi}e^{kL}$.
The use of the physical localizations here helps us understand the mass scales associated with the couplings of different fields.
Despite the zero-mode scalar being massless in 5D and having couplings which are independent of the $y$-coordinate, the typical scale of its interactions are determined by the physical localization.
Just as the KK graviton in Randall-Sundrum models has couplings which are independent of the $y$ coordinate, and solves the gauge-hierarchy problem by having a physical wave-function localization in the UV end of the extra dimension, here too the zero-mode scalar has suppressed interactions because its physical wave-function is peaked at $y=L$.

In \rcite{Giudice:2017suc} it was proposed that the following form of brane interaction could be used in the linear dilaton model,
\beq
\tilde{S}_{\rm int}=-\int \!d^4x\int_0^L\!dy\delta(y-\hat y)\left(Y\frac{e^{ky}\pi}{f_{\pi}}\bar{\psi}\psi\right)
\eeq
with the exponential factor multiplying the brane interaction term. However, as pointed out in \rcite{Giudice:2017suc}, such an interaction originating from a linear dilaton coupling would lead to explicit breaking of conformal invariance of the background geometry.
In this case once we canonically normalize the Goldstone and the fermion kinetic terms we arrive at the following interactions
\beq
\tilde{\mathcal{L}}_{\rm int}= -\frac{N_0~Y}{f_{\pi}e^{k(L-\hat y)}}\pi_0(x)\bar{\psi}\psi -\frac{Yc_n}{f_\pi}\pi_n(x)\bar{\psi}\psi.
\eeq
Note that in such scenarios, which are rather ad-hoc in the 5D setup, the zero-mode couplings are $\hat{y}$-dependent, whereas the couplings of the gears are independent of $\hat y$. 
We will not discuss such model building possibilities any further since the basic properties of the clockwork mechanism do not require 5D position dependent couplings for the zero-mode.

\subsubsection*{Relation to the discrete model} \label{5D4D}

In taking the continuum limit of the discrete clockwork model we take $N\rightarrow\infty$, however at the same time we must require $q\rightarrow1$ such that $q^N$ remains constant.
In order to make a connection between the 5D and 4D discrete models we must match some of the observables to get a relation between the 4D and 5D parameters.
Fixing $q^N=W$ (constant) and matching the masses obtained from the 5D approach to those from the discrete model (Eqs. \ref{cweq3} and \ref{mass_sq}) we find
\begin{align}
L=\frac{N+1}{\epsilon f_{\pi}},	\lsp \text{and} \lsp
k\simeq\frac{\ln(W)}{N}\epsilon f_{\pi}.
\end{align}
which implies that $kL\!\simeq\!\ln(W)$ and thus $q^N\!\simeq\!e^{kL}$.
In the discrete case we have seen that the Goldstone boson had $q^{-N}$ suppressed couplings to fields which coupled to the $N^{\text{th}}$ site of the model.  Now we see that this is identical to the couplings of the zero-mode Goldstone boson described in the 5D model.
Of course this was pointed out already in \cite{Giudice:2016yja}, here we have simply presented a 5D model that naturally gives us a massless 5D scalar field with the same global symmetries as the discrete case.
At the same time we have shown that the radial fluctuations can be decoupled in a way which is entirely dependent on the mass gap feature of the clockwork models.
In this paper we discuss more general scenarios in which we can have couplings in the 5D model that are not restricted to the $y\!=\!0$ brane, however in doing so we make some departures from the discrete model described in section \ref{clockwork}.

The higher order interactions between the gears do not survive in the continuum limit since they scale inversely with $N$.
This would even be true of interactions involving the zero-mode (which is relevant to the non-abelian cases), since its normalisation constant in the continuum limit becomes
\beq  \label{N0cont}
\mathcal{N}_0\simeq\sqrt{\frac{2\ln(W)}{N}}.
\eeq
Although the higher order interactions in the discrete model do not survive in the continuum limit, interactions involving more fields are expected when we add higher derivative terms to the 5D model.
The higher derivative terms in the 5D action are built from factors of $\partial_MU$, and due to the flat 5D profile of the zero-mode longitudinal component we find that no non-derivative interactions for these modes are present in the effective theory, as expected.
Non-derivative interactions for the KK modes are however present, just as they are for the gears in the discrete model.
In appendix \ref{appendixA} we study these terms in some detail~\footnote{A related study has also been recently presented in \cite{Ben-Dayan:2017rvr}.}, and show that only in some cases do these interactions have a parity-like discrete symmetry similar to that found in Eq.~\eqref{parity2} for the discrete clockwork scenario.

\subsection{A clockwork scalar from a 5D gauge field}
\label{abelianA5}

Models employing a clockwork mechanism to generate hierarchically small couplings for gauge fields have been discussed already in the literature \cite{Giudice:2016yja}.
And the continuum limit of these models can be built from a bulk 5D gauge field in the LD model.
In this section we study the scalar component of the abelian bulk gauge field $A_5$.
We start with the following action in the Jordan frame of the LD background geometry~\footnote{In Appendix~\ref{gauge_LDM} we give a detailed description of the abelian gauge model in the LD background geometry both in the Jordan and Einstein frames.},
\beq
S_A=\int\! d^4x\int_0^L\!dy~e^{2ky}\left[-\frac{1}{4}F^{MN}F_{MN}+\frac{1}{2\xi}\left(\partial_{\mu}A^{\mu}+\xi e^{-2ky}\partial_5(e^{2ky}A_5)\right)^2\right],   \label{5Dgaugeac}
\eeq 
where the gauge fixing term cancels the kinetic mixing between the $A_\mu$ and $A_5$ modes.
After performing Kaluza Klein decompositions we find that the massive $A_5$ modes are eaten by the massive $A_{\mu}$ modes, providing them with longitudinal degrees of freedom in the unitary gauge ($\xi\to\infty$).
Then, depending on boundary conditions, the $A_{\mu}$ and $A_5$ modes can have massless zero-mode solutions.
If we want to have only a zero-mode gauge field we would impose
\beq
\partial_5A_{\mu}\Big|_{y=0,L}=0,\hsp\text{and}\hsp A_5\Big|_{y=0,L}=0.	\label{n_bc}
\eeq
And if we want to have only a zero-mode $A_5$ scalar we impose
\beq
A_{\mu}\Big|_{y=0,L}=0,	\hsp \text{and}\hsp \partial_5\big[e^{2ky}A_5\big]\Big|_{y=0,L}=0.	\label{d_bc}
\eeq	
It has been shown in \cite{Giudice:2016yja} that if the zero-mode gauge field solution is allowed then the zero-mode gauge boson shares key properties with the clockwork gauge bosons in the discrete models, i.e. the suppressed gauge coupling and distinct mass spectra of gears.
However in this work we will focus on the scenario with an $A_5$ scalar zero-mode and no zero-mode $A_\mu$ gauge boson.
Therefore we will have a model a spin-0 zero-mode and spin-1 gears. Performing a KK decomposition on these $A_\mu$ and $A_5$ fields as 
\beq
A_{\mu(5)}(x^\mu,y)= \frac{e^{-ky}}{\sqrt{L}} \sum_{n=0}^\infty A_{\mu(5),n}(x^\mu)
f^{A(5)}_n(y)~,
\label{gaugeKK}
\eeq
and evaluating the eigenvalue equation to determine the 5D profiles (details are given in the Appendix~\ref{gauge_LDM}). 

Applying Dirichlet boundary conditions on the gauge KK wave-functions in the Feynman gauge, we find
\beq
f^A_n(y)=\sqrt2\sin\left(\frac{n\pi y}{L}\right), \lsp{\rm with}\hsp m_{A,n}^{2}=k^2+\left(\frac{n\pi}{L}\right)^2.
\eeq
As we can see, the spin-1 KK modes of this scalar clockwork model have the same mass spectrum as the spin-0 KK modes in the previous section.
From the action and the boundary condition \eqref{d_bc} we can see that the zero-mode of $A_5$ will have an exponentially localized zero-mode solution
\beq
f^5_{0}(y)=N_0~e^{-ky}, \lsp{\rm where}\hsp N_0\equiv \sqrt{\tfrac{2kL}{1-e^{-2kL}}}~.
\eeq
Whereas, the KK-mode wave-functions for the $A_5$ field are fixed by \eqref{A5profile}, i.e.
\beq
f^5_{n}(y) = \frac{e^{ky}}{m_{A,n}} \partial_5\big[e^{-ky} f^A_{n}(y)\big]= \frac{\sqrt2 \,n\pi}{m_{A,n}L}\cos\left(\frac{n\pi y}{L}\right).	\label{A5KKmodes}
\eeq
It is important to mention here that it is not surprising that the $A_5$ KK wave-functions are determined by the $A_\mu$ KK wave-functions since this is a manifestation of the KK Higgs mechanism in an extra dimension, where in {\it unitary gauge} $(\xi\!\to\!\infty)$ the KK-modes $A_{\mu,n}$ would eat the KK-modes of $A_{5,n}$ at each KK level and become massive. However, in order to establish a duality with the 4D discrete model of this 5D model, we work in the $R_\xi$ gauge in the following subsection.

The exponential localization of this zero-mode gauge scalar, $A_{5,0}$, towards the $y\!=\!0$ brane makes the couplings of this mode significantly different than the massless Goldstone boson from the previous section.
To see this we will consider interactions of the bulk gauge field, $A_M$, and a scalar field living on a brane at some position $\hat y$ in the extra dimension, 
\beq
S_{\rm int}^A=-\int \!d^4x\int_0^L\!dy\delta(y-\hat y)\Big[g_5^2 A_M^2|\phi|^2\Big].
\eeq
We find the following position dependent couplings in the effective theory for the zero-mode scalar $A_{5,0}$ and KK-mode of the gauge field $A_{\mu,n}$,
\beq
\mathcal{L}_{\rm int}^A=-g_4^2~e^{-2k\hat y}\Big(e^{-2k\hat y}A_{5,0}^2|\phi|^2-c_n A_{\mu,n}^2|\phi|^2\Big),
\eeq
where $g_4\!\simeq \!\sqrt{2k}\,g_5$ and $c_n\!\sim\!\op(1)$. If we consider the interactions at $\hat y\!=\!L$, then the scalar zero-mode coupling is exponentially suppressed in comparison to the couplings of the massive gauge gears. 
The reason for this can be understood from the fact that the physical localization of the $A_5$ zero-mode is exactly opposite to that of the would be zero-mode gauge field, and therefore its couplings scale in a different way to the couplings of the KK gears.

Now it is interesting to consider the $k\!<\!0$ case.  Here we see that the physical localizations all flip and the normalization constant of the $A_5$ zero-mode grows exponentially large.  Thus the $A_5$ zero-mode can indeed exhibit clockwork suppressed couplings.
If we were to do the same in the bulk scalar case we would see that the zero-mode normalization constant becomes $\mathcal{O}(1)$ and its physical localization shifts to the $y\!=\!0$ brane, with the same thing happening for the $A_{\mu}$ zero-mode of a bulk gauge field.
The reasons for this will become clearer in the discrete 4D analogue of this model presented in section \ref{discreteA5}.

\subsubsection{Deconstruction}

\noindent This abelian $A_5$ model can be deconstructed following \cite{ArkaniHamed:2001ca,ArkaniHamed:2001nc,Hill:2000mu}. In doing so we arrive at an effective theory for $N\!+\! 1$ Goldstone bosons and $N$ massive gauge fields.
Let us start by rescaling the 5D fields $A_{\mu,5}\to e^{-ky}A_{\mu,5}$ to get the canonical form of the 4D kinetic terms of Eq.~\eqref{5Dgaugeac}, this results in
\begin{align}
S=\int d^4x\int_0^Ldy\Big[&-\frac{1}{2}\left(\partial^{\mu}A^{\nu}\partial_{\mu}A_{\nu}-\partial^{\mu}A^{\nu}\partial_{\nu}A_{\mu}\right)+\frac{1}{2}e^{2ky}(\partial_ye^{-ky}A_{\mu})^2+\frac{1}{\xi}(\partial_{\mu}A^{\mu})^2 \non\\
&+\frac{1}{2}(\partial_{\mu}A_5)^2+\frac{\xi}{2}e^{2ky}(\partial_ye^{-ky}A_5)^2\Big]. 
\end{align}
Let us discretize the extra dimension by replacing $y=j\ell$ and $N\ell=L$, where $\ell$ is the length of the step size on the discrete lattice and $N+1$ are the total number of lattice sites. In the discrete version the integral and the derivatives are replaced as, 
\beq
\int_0^Ldy \to \sum_{j=0}^{N-1} \ell, \Lsp \frac{df(y)}{dy}\to \frac{f_{j+1}(\ell)-f_{j}(0)}{\ell}.
\eeq
With these observations one can rewrite the above action in the discretized theory as, 
\begin{align}
S=\int d^4x\bigg\{\sum_{j=0}^N&\left[-\frac14F^j_{\mu\nu}F_j^{\mu\nu}+\frac1\xi \big(\partial_\mu A^j_\mu\big)^2+\frac12\big(\partial_\mu A^j_5\big)^2\right]	\notag\\
&+\frac1{\ell^2}\sum_{j=0}^{N-1}\big(A^j_\mu-e^{-k\ell}A^{j+1}_\mu\big)^2+\frac1{\ell^2}\frac\xi2\sum_{j=0}^{N-1}\big(A^j_5-e^{-k\ell}A^{j+1}_5\big)^2\bigg\}.
\end{align}
Since the boundary conditions for the $A_\mu$ field are Dirichlet, i.e. $A_\mu(y\!=\!0)=0$ and $A_\mu(y\!=\!L)=0$, which translate in the discretize theory as, 
$A^{j\!=\!0}_\mu=0$ and $A^{j\!=\!N}_\mu=0$.
Hence the discrete action takes the form, 
\begin{align}
S=\int d^4x\bigg\{\sum_{j=1}^{N-1}&\left[-\frac14F^j_{\mu\nu}F_j^{\mu\nu}+\frac1{2\xi} \big(\partial_\mu A^j_\mu\big)^2\right]+\sum_{j=0}^{N}\frac12\big(\partial_\mu A^j_5\big)^2	\notag\\
&+m_A^2\sum_{j=1}^{N-1}\big(qA^j_\mu-A^{j+1}_\mu\big)^2+\frac\xi2m_A^2\sum_{j=0}^{N-1}\big(A^j_5-qA^{j+1}_5\big)^2\bigg\}.
\end{align}
where we defined $q\equiv e^{k\ell}$ and $m_A^2\equiv 1/\ell^2$.
Diagonalising these mass mixings results in a massless $A_5$ mode and massive vector modes just like in the continuum case.
Again the $A_{\mu}$ modes eat an asymmetric combination of the $A_5$ modes leaving the true $A_5$ Goldstone boson with an asymmetric localisation along the lattice.
In the next subsection we will discuss a 4D model consisting of scalars and gauge fields in which the symmetry breaking pattern is chosen to reproduce the physics of this deconstructed model.

\subsubsection{Discrete gauged clockwork scalar} \label{discreteA5}
\begin{figure}[t]
\centering
\includegraphics[width=\textwidth]{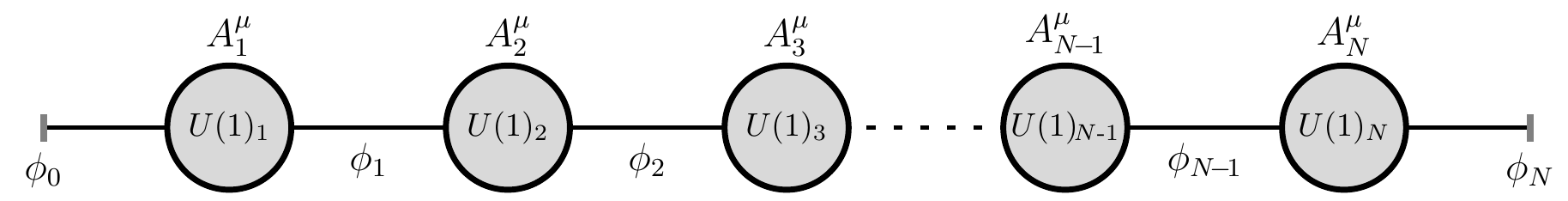}
\caption{Illustration of the $N$-site discrete abelian gauge model, where $\phi_j$ is a complex link scalar field between the adjacent gauge groups.}
\label{fig:cgbA5}
\end{figure}

We take an $N\!+\!2$-site discrete model consisting of $U(1)_j$ gauge groups with gauge field $A_\mu^j$ ($j\!=\!0,\ldots,N\!+\!1$) and universal gauge coupling $g$ at each site. Now we link the $N\!+\!2$-site gauge theory with $N\!+\!1$ complex scalar fields $\phi_j$ which have charges $(1,-\!q)$ under adjacent gauge groups $U(1)_j\!\times\!U(1)_{j+1}$. In the 5D model we impose Dirichlet boundary conditions for the $A_{\mu}$ field at each brane in the extra dimension.
In this 4D deconstruction this corresponds to
$A^{j\!=\!0}_\mu=0$ and $A^{j\!=\!N\!+\!1}_\mu=0$, i.e. the boundary sites are not gauged.  Therefore we have $N\!+\!1$ scalar fields and just $N$ gauge fields, as sketched in Fig.~\ref{fig:cgbA5}.

Now we can write the discrete action for the model described above with $N$ gauge fields and $N\!+\!1$ complex scalar fields as
\beq \label{axionpotential2}
S=\int d^4x\bigg\{\sum_{j=1}^{N}\Big[-\frac14F^j_{\mu\nu}F_j^{\mu\nu}\Big]+\sum_{j=0}^{N}\Big[D_{\mu}\phi_j^\dag D^{\mu}\phi_j-\mu^2\phi_j^{\dagger}\phi_j-\frac{\lambda}{4}\vert\phi_j^{\dagger}\phi_j\vert^2\Big]\bigg\}.
\eeq
The covariant derivatives are chosen such that above action is invariant under the following transformations of the complex scalar fields,
\begin{align}
\phi_0\rightarrow& e^{-iq\alpha_1(x)}\phi_0,\non \\
\phi_j\rightarrow& e^{i(\alpha_{j}(x)-q\alpha_{j+1}(x))}\phi_j, \Lsp (j=1\ldots,N\!-\!1) \\
\phi_N\rightarrow& e^{i\alpha_N(x)}\phi_N, \non
\end{align}
with the gauge fields transforming as 
 \beq
 A^{j}_{\mu}\rightarrow A^{j}_{\mu}-\frac{1}{g}\partial_{\mu}\alpha_j(x).
 \eeq
 The covariant derivatives are then
 \begin{align}
 D_{\mu}\phi_0\equiv& \left(\partial_{\mu}-igqA^{1}_{\mu}\right)\phi_0 \non\\
 D_{\mu}\phi_j\equiv& \left(\partial_{\mu}+igA^{j}_{\mu}-igqA^{j+1}_{\mu}\right)\phi_j,\lsp (j=1,\ldots,N-1)  \\
 D_{\mu}\phi_N\equiv& \left(\partial_{\mu}+igA^{N}_{\mu}\right)\phi_N.	\non
 \end{align}
Choosing a scalar potential such that the $\phi_j$ fields acquire vacuum expectation values, i.e. $\langle\phi_j\rangle\!=\!f_{\pi}\!=\!\sqrt{-2\mu^2/\lambda}$. The radial modes have masses of the order of $\sqrt2 f_\pi$ and taking $g\ll1$ they can be integrated out at the mass scale of the vector modes ($\sim\! g f_{\pi}$). Below the scale $f_\pi$ we can use a gauged non-linear sigma model and we see that the mass matrix for the gauge bosons in now similar to that of Eq.~\eqref{clockworkMassM} but with additional contributions on the $j\!=\!1$ and $j\!=\!N$ sites.

We can rewrite the above discrete action in the $R_\xi$ gauge as, 
\begin{align}
S=\int d^4x\bigg\{\sum_{j=1}^{N}&\left[-\frac14F^j_{\mu\nu}F_j^{\mu\nu}+\frac1{2\xi} \big(\partial_\mu A^j_\mu\big)^2\right]+\sum_{j=0}^{N}\frac12\big(\partial_\mu \pi_j\big)^2	\notag\\
&+\frac12m_A^2\sum_{j=0}^{N}\big(A^j_\mu-qA^{j+1}_\mu\big)^2+\frac\xi2 m_A^2\sum_{j=0}^{N-1}\big(q\pi_j-\pi_{j+1}\big)^2\bigg\}.
\end{align}
where we defined $m_A^2\equiv g^2f_\pi^2$ and remembering that the Dirichlet boundary conditions are in force, $A^0_\mu\!=\!A^{N\!+\!1}_\mu\!=\!0$.
Now we write the discrete site Lagrangian in the following form, 
\begin{align}
\cL&=\sum_{j=1}^{N} \left[-\frac14F^j_{\mu\nu}F_j^{\mu\nu}+\frac1{2\xi} \big(\partial_\mu A^j_\mu\big)^2\right]+\sum_{j=0}^{N}\frac12\big(\partial_\mu \pi_j\big)^2	- \frac12\sum_{i,j=1}^{N}A^i_\mu{\cal M}^2_{A,ij}A^{j}_\mu-\frac12\sum_{i,j=0}^{N}\pi_i{\cal M}^2_{\pi,ij}\pi_j,		\label{lagphi_nsite}
\end{align}
where the mass matrices for the vectors $\vec A_\mu\equiv(A^1,A^2,\cdots,A^{N})_\mu$ and $\vec \pi\equiv(\pi^0,\pi^1,\cdots,\pi^N)$ are dimensions $N\!\times\! N$ and $(N\!+\!1)\!\times\!(N\!+\!1)$, respectively, and are given as, 
\begin{align}
{\cal M}^2_{A,ij}&=m_A^2
\bpm
1+q^2 & -q & \cdots & 0 & 0 \cr 
-q & 1+q^2 & \cdots & 0& 0 \cr
\vdots &  \vdots & \ddots &\vdots &\vdots \cr
 0&0 &  \cdots& 1+q^2 & -q \cr
 0 & 0 &\cdots & -q & 1+q^2
\epm_{N\times N}, \label{AmumassM}\\
{\cal M}^2_{\pi,ij}&=\xi m_A^2
\bpm
q^2 & -q & 0 & \cdots & 0 & 0 \cr 
-q & 1+q^2 & -q & \cdots & 0& 0 \cr
0 & -q & 1+q^2 & \cdots & 0& 0 \cr
\vdots & \vdots & \vdots & \ddots &\vdots &\vdots \cr
 0&0 & 0& \cdots& 1+q^2 & -q \cr
 0 & 0 & 0 &\cdots & -q & 1
\epm_{(N\!+\!1)\times(N\!+\!1)}.
\label{pimassM}
 \end{align}
Note that mass matrix ${\cal M}^2_{\pi,ij}$ for the $\vec \pi$ fields \eqref{pimassM} is same as for the clockwork scalar in Eq. \eqref{clockworkMassM} except the first and last entries are reversed, as a result the orthogonal rotation for zero-mode $\op_{j0}$ would be, 
\beq
\op_{j0}={\cal N}_0q^j, \lsp {\rm where } \lsp{\cal N}_0\sim\op(1),
\eeq
however the non-zero mass matrix $\op_{jk}$ has the same form as given in Eq.~\eqref{ClockworkRot} with the mass eigenvalues, 
\beq \label{mass_sq_xi}
m_{\pi^\p,k}^2=\xi\lambda_k g^2f_\pi^2, \lsp\text{for}\hsp k=1,\ldots,N,
\eeq
where $\lambda_k$ is defined in Eq.~\eqref{cweq3}. It is important to note that the massive `gears' for the $\pi$ field are gauge ($\xi$) dependent, which in the unitary gauge $\xi\to\infty$ can be gauged away from the spectrum. Hence leaving us with only zero-mode which is asymmetrically distributed among all the $\pi$ fields.  
From the results of the previous section we entirely expected this mass matrix to have the $j\!=\!0,N$ sites flipped with respect to the mass matrix in Eq.~\ref{clockworkMassM}.
This reflects the difference in physical localizations between the $A_5$ zero-mode and the zero-mode Goldstone arising in the spontaneous symmetry breaking of a bulk scalar potential.
It is important to note that this discrete model presented here is, in its own right, a fully UV complete 4D clockwork theory in which fields coupled to the $N^{\text{th}}$ site have clockwork suppressed couplings to the true Goldstone when $q\!<\!1$.

Now let us turn our attention to diagonalize the ${\cal M}^2_{A,ij}$ for the $\vec A_\mu$. We find the following orthogonal transformation diagonalizing the $A_\mu$ mass matrix,
\beq \label{ClockworkRotAmu}
\mathcal{O}^{A}_{jk}=\mathcal{N}_k\sin\frac{jk\pi}{N\!+\!1},\hsp\text{with}\hsp j,k=1,\ldots,N,
\eeq
where $\mathcal{N}_k$ is defined in Eq.~\eqref{cweq3}. An important observation is that there is no normalisable zero-mode in the spectrum of gauge fields. All the gauge fields $A_\mu$ are massive with masses, 
\beq \label{mass_sq_A}
m_{A^\p,k}^2=\lambda_kg^2f_\pi^2, \lsp\text{for}\hsp k=1,\ldots,N.
\eeq
As in the 5D case, there is a tower of massive KK-modes for the $A_{\mu,n}$ fields and a massless $A_{5,0}$ scalar asymmetrically distributed along the lattice.
We have shown already how the scalar mass spectrum in the discrete model can be matched to the KK spectrum of the 5D theory, and the exact same matching can be done here.  Thus we have demonstrated that the scalar $A_5$ zero-mode of a bulk gauge field in the linear dilaton model exhibits a clockwork mechanism, and have described a consistent UV complete 4D deconstruction of this model built from scalar fields and $U(1)$ gauge fields.
Moreover, this UV complete discrete model involves no explicit symmetry breaking whatsoever, and all couplings are fixed by gauge invariance.

\section{Non-abelian clockwork}
\label{Non-abelian clockwork}

In this section we explore a possible non-abelian extension of the abelian clockwork model described in \cite{Giudice:2016yja} and recapped section \ref{clockwork} of this paper.
The basic set-up includes a non-abelian global symmetry ($\mathcal{G}_j$) at each site, spontaneously broken to a subgroup $\mathcal{H}_j$.
A clockwork interaction, analogous to that in the abelian case, then generates mixing between the Goldstone bosons at neighboring sites.
Diagonalizing these mixings we find a massless zero mode whose interactions to fields external to the clockwork sector can be hierarchically suppressed in the same way as in the abelian case.
However we will see below that the clockwork interaction required to generate the exponential hierarchy explicitly breaks the global symmetries, as can be seen by examining the quartic terms in the potential.

Suppose that we have a chain with $N+1$ sites labelled $j=0,\ldots,N$, and at each site we have a scalar multiplet transforming under the global symmetry $\mathcal{G}_j=SU(\cN)_{L}^j\times SU(\cN)_{R}^j$, such that
\beq
\Phi_j\rightarrow L_j\Phi_jR_j^{\dagger}\,,
\eeq
where $L_j$ and $R_j$ are global transformations of $SU(\cN)^j_L$ and $SU(\cN)^j_R$ symmetries, respectively. 
The action can be written as
\beq \label{NAaction}
\mathcal{L}=\sum_{j=0}^N\Big[\text{Tr}\big(\partial_{\mu}\Phi^\dagger_j\partial^{\mu}\Phi_j\big)-\mu^2\text{Tr}\big(\Phi_j^\dagger\Phi_j\big)-\frac{\lambda}{4}\text{Tr}\Big(\big(\Phi^\dagger_j\Phi_j\big)^2\Big)\Big]
\eeq
such that each $\Phi_j$ acquires a VEV $f_{\pi}$ for $\mu^2\!<\!0$.
This spontaneously breaks the global symmetries at each site to the vectorial subgroup $\mathcal{H}_j=SU(\cN)^j_V$, i.e. $L_j\!=\!R_j$.
The clockwork mechanism in the abelian case was generated by the interaction term in Eq.~\eqref{axionpotential}.
Here, at a scale $\epsilon f_{\pi}\ll f_{\pi}$, we introduce an analogous interaction
\begin{align}\label{cw1}
\delta\mathcal{L}=\epsilon^2f_{\pi}^{3-q}\sum_{j=0}^{N-1}\text{Tr}\Big(\Phi_j^{\dag}\Phi_{j+1}^q+\text{h.c.}\Big),
\end{align}
where again $q\!>\!1$ and is an integer~\footnote{However, this condition that $q\!\in\!\mathbb{Z}$ could be relaxed if we only consider an EFT description with a non-linear sigma model as we do later in this work. In such a case there exists a continuum limit which could be described as a 5D linear dilaton model, as is also discussed in the previous sections for the abelian cases.}.
This term introduces an explicit breaking which respects just one linear combination of the global symmetries, this diagonal combination is given by $L_j\!=\!R_j\!=\!L_{j+1}\!=\!R_{j+1}$ and we denote it as $\mathcal{H}^{\prime}=SU(\cN)^{\prime}_V$.
In the limit $q\rightarrow1$ a larger symmetry group is preserved where $L_j\!=\!L_{j+1}$ and $R_j\!=\!R_{j+1}$ independently, this we denote $\mathcal{G}^{\prime}=SU(\cN)^{\prime}_L\times SU(\cN)^{\prime}_R$. 
Thus when $q\!\neq\!1$ the $G^{\prime}$ symmetry is explicitly broken in such a way that preserves only the diagonal subgroup $\mathcal{H}^{\prime}$.
This explicit breaking is a crucial difference between the abelian and non-abelian clockwork models.

Assuming that the scale of these explicit clockwork breaking is much lower than the scale at which the vector-like symmetries are broken to their diagonal subgroups, we can describe the dynamics of the $N+1$ (pseudo-)Goldstone boson multiplets using a non-linear sigma description where the singlet fluctuation has been integrated out,
\beq
\delta\mathcal{L}_{\rm EFT}=\epsilon^2f_{\pi}^4\sum_{j=0}^{N-1}\text{Tr}\left(U_j^{\dagger}U_{j+1}^q+\text{h.c.}\right). 	\label{deltaL_EFT}
\eeq
We have taken
\beq \label{Phi}
\Phi=\Sigma_VU_j\,, \lsp \text{with}\hsp U_j=e^{-i\Pi_j/f_{\pi}}\,,~~{\rm and}~~ \Pi_j\equiv\pi_j^{\hat{a}}(x)T^{\hat{a}}\,,
\eeq
where $T^{\hat{a}}$ are the broken generators normalized via $\text{Tr}[T^aT^b]=\tfrac{1}{2}\delta^{ab}$ in the fundamental representation, $\pi_j^{\hat{a}}(x)$ are the (pseudo-)Goldstone degrees of freedom, and $\Sigma_V$ is the VEV breaking the global symmetry. 
Note that in what follows below we are only interested in the EFT description of this non-abelian clockwork model, thus we work in the non-linear sigma model framework \eqref{deltaL_EFT} and as a consequence allow $q\!\in\! \mathbb{R}$.
Expanding the potential up to quartic order in fields, we find
\begin{align}\label{SUNpot}
V(\Pi)=\sum_{j=0}^{N-1}\Big[\epsilon^2f_{\pi}^2\text{Tr}\left((\Pi_j-q\Pi_{j+1})^2\right)-\frac{\epsilon^2}{6}\text{Tr}\left((\Pi_j-q\Pi_{j+1})^4\right)\nonumber \\
-\frac{\epsilon^2q^2}{3}\text{Tr}\left(\Pi_j[\Pi_j,\Pi_{j+1}]\Pi_{j+1}\right)\,+\,\text{h.c.}\Big].
\end{align}
Terms on the first line are analogous to those appearing in the abelian case, however the terms on the second line appear due to the non-commutativity of the generators.
In the mass eigenbasis the above potential becomes
\begin{align}  \label{PotEq}
V(\Pi^{\prime})=\sum_{klmn}\Big[&\frac{m_k^2}{2}\pi_k^{\hat{a}\prime}\pi_k^{\hat{a}\prime}-\frac{\epsilon^2}{6}\mathcal{C}_{klmn}\text{Tr}\left(T^{\hat{a}}T^{\hat{b}}T^{\hat{c}}T^{\hat{d}}\right)\pi_k^{\hat{a}\prime}\pi_l^{\hat{b}\prime}\pi_m^{\hat{c}\prime}\pi_n^{\hat{d}\prime}    \non\\
&-\frac{\epsilon^2q^2}{3}\mathcal{W}_{klmn}\text{Tr}\left(T^{\hat{a}}[T^{\hat{b}},T^{\hat{c}}]T^{\hat{d}}\right)\pi_k^{\hat{a}\prime}\pi_l^{\hat{b}\prime}\pi_m^{\hat{c}\prime}\pi_n^{\hat{d}\prime}\,+\,\text{h.c.} \Big],
\end{align}
where $m_k^2$ and $\mathcal{C}_{klmn}$ are the same parameters defined in Eqs.~\eqref{mass_sq} and \eqref{Cklmn} for the abelian case and so do not contain interactions for the zero mode scalars.
The $\mathcal{W}_{klmn}$ term however is written as
\beq \label{Weq}
\mathcal{W}_{klmn}=\sum_{j=0}^{N-1}\mathcal{O}_{jk}\mathcal{O}_{jl}\mathcal{O}_{j+1,m}\mathcal{O}_{j+1,n}.
\eeq
We see that this term can be non-zero for configurations involving the zero modes, therefore the $\Pi_0^{\prime}$ fields are pseudo-Goldstone bosons rather than being exact Goldstone bosons, i.e. they are massless at the tree-level however their quartic interactions suggest that at the quantum level a mass will be generated.
This is a direct consequence of the fact that the interaction terms in Eq.~\eqref{cw1} generally only preserve $\mathcal{H}^{\prime}$, with the explicit breaking of $\mathcal{G}^{\prime}$ leading to interactions for the zero mode, $\Pi_0^{\prime}$.

Up until now the simplifications we have made are general to any global symmetry structure, not just the product group we have chosen here.
It is only when we evaluate the traces that the model dependence enters.
The remaining structure of the quartic interactions are determined by the traces, which we can evaluate for the $\mathcal{W}$ term as
\begin{align}
\text{Tr}\Big(T^{\hat{a}}[T^{\hat{b}},T^{\hat{c}}]T^{\hat{d}}\Big)&=if_{\hat b\hat c\hat e}\text{Tr}\Big(T^{\hat{a}}T^{\hat{e}}T^{\hat{d}}\Big) \non\\
&=\frac{i}{4}f_{\hat b\hat c\hat e}\Big(d_{\hat a\hat e\hat d}+if_{\hat a\hat e\hat d}\Big)	\non\\
\text{Tr}\Big(T^{\hat{a}}[T^{\hat{b}},T^{\hat{c}}]T^{\hat{d}}\Big)\,+\,\text{h.c.}&=\tfrac{1}{2}f_{\hat b\hat c\hat e}f_{\hat a\hat d\hat e}
\end{align}
where $f_{\hat a\hat b\hat c}$ and $d_{\hat a\hat b\hat c}$ are the antisymmetric structure constants and symmetric constants of the $SU(\cN)$ algebra, respectively.
For the trace over the generators in the $\mathcal{C}$ term we have
\begin{align} 
\text{Tr}\Big( T^{a}T^{b}T^{c}T^{d} \Big) =& \frac{1}{4\cN}\big( \delta_{ab}\delta_{cd} + \delta_{ad}\delta_{bc} - \delta_{ac}\delta_{bd} \big) \non\\ 
& + \frac{1}{8} \big( d_{abe}d_{cde} + d_{ade}d_{bce} - d_{ace}d_{bde} \big) \non\\ & + \frac{i}{8} \big( d_{abe}f_{cde} + d_{cde}f_{abe} \big)
\end{align}
with the imaginary term vanishing under the hermitian conjugate.
Further simplifications of the potential here are not straightforward, so we will refrain from presenting lengthy expressions.
The purpose of this section is a demonstration of how the higher order terms from the potential do not respect the global symmetries $\mathcal{G}_j$.
We know from Eqs.~\eqref{parity} and \eqref{N0cont} that as $N\to\infty$ the higher order terms that break these global symmetries go to zero.  Therefore in the continuum limit we should expect to be able to build EFT of non-abelian clockwork models which fully respect the global symmetries.

\subsection{non-abelian clockwork from a 5D scalar}

In this section we generalize the continuum limit of the abelian clockwork model described in the previous section to non-abelian global symmetries. 
As a simple example we again consider $SU(\cN)_L\times SU(\cN)_R$ and for concreteness we limit ourselves to $\cN\!=\!2$, implementing this as a global symmetry in the 5D bulk of the LD model. 
We consider a scalar multiplet $\Phi(x,y)$ which transforms under this bulk global symmetry as $\Phi\to L\Phi R^\dag $ where $L$ and $R$ represent the global transformations of $SU(2)_L$ and $SU(2)_R$ global symmetries, respectively. 
The 5D action for such a theory can be written as, 
\beq \label{5DNaction}
S_{\textsc{na}}= \int\! d^4x\int_0^L\!dy~e^{2ky}\Big\{\text{Tr}\big[\partial_{M}\Phi^\dagger\partial^{M}\Phi\big]-\mu^2\text{Tr}\big[\Phi^\dagger\Phi\big]-\frac{\lambda}{4}\text{Tr}\big[\big(\Phi^\dagger\Phi\big)^2\big]\Big\}\,,
\eeq 
where for the sake of brevity we use the same notation of $\mu$ and $\lambda$ as in the abelian case. 
Analogous to the abelian case we do not include brane potentials, and we thus find that when $\mu^2\!<\!0$ the field $\Phi(x,y)$ obtains a VEV given by $\Sigma_V=\mathbb{I}\sqrt{\tfrac{2\mu^2}{\lambda}}=\mathbb{I}\tfrac{v}{\sqrt L}$, such that $[v]\!=\!1$.
The 5D profile of this VEV is again flat due to the fact that we have no brane potentials for the scalar field.
The vectorial subgroup left unbroken in the vacuum is denoted by $SU(2)_V$ whereas the broken axial subgroup is denoted $SU(2)_A$.

The field $\Phi(x,y)$ has four degrees of freedom, three of which are identified as Goldstone bosons of the global symmetry breaking, and another which is identified with the radial fluctuation.
We can write this field as
\beq
\Phi(x,y)=\tfrac{1}{\sqrt{2L}}\left(v+r(x,y)\right)e^{i\sqrt{2}T_A^{\hat{a}}\pi^{\hat{a}}(x,y)/v}
\eeq
where $r(x,y)$ is the radial fluctuation and $\pi^{\hat{a}}(x,y)$ is the Goldstone fluctuation along the broken generator $T_A^{\hat{a}}$, and have mass dimensions $[r(x,y)]\!=\![\pi^{\hat a}(x,y)]\!=\!1$.
Following the same procedure as in the abelian case and decomposing these fields in a KK decomposition as,
\beq
r(x,y)=e^{-ky}\sum_{n=0}^{\infty}r_n(x)f^{r}_n(y), \lsp \pi^{\hat a}(x,y)=e^{-ky}\sum_{n=0}^{\infty}\pi^{\hat a}_n(x)f^{\pi}_n(y),
\eeq
we find that the 5D profiles are the same as in the abelian case, i.e.
\beq
f^{r(\pi)}_n(y)=\frac{\sqrt2 \, n\pi}{m_{r(\pi),n}L}\Big[\cos\left(\frac{n\pi y}{L}\right)+\frac{kL}{n\pi}\sin\left(\frac{n\pi y}{L}\right)\Big].
\eeq
where the mass spectrum of these components are also analogous to those in the abelian case, i.e.
\begin{align}
r_n(x):&\lsp  m_{r,n}^2=k^2+2\mu^2+\left(\tfrac{n\pi}{L}\right)^2\,,  \nonumber \\
\pi^{\hat{a}}_n(x):&\lsp m_{\pi,n}^2=k^2+\left(\tfrac{n\pi}{L}\right)^2\,.
\end{align}
The axial components of the 5D scalar field contain massless zero modes in their KK towers, these fields are the Goldstone bosons of the spontaneous symmetry breaking.
The profile of these canonically normalized zero modes $\pi^{\hat a}_0(x)$ is simply
\beq
f^{\pi}_0(y)=N_0~e^{k(y-L)},	\lsp{\rm where} \hsp N_0\equiv \sqrt{\tfrac{2kL}{1-e^{-2kL}}}\,,
\eeq
again matching that in the abelian case.
Assuming that $k^2\ll\mu^2$ we can take an effective theory with just the axial modes, resulting in exactly the theory one desires from a non-abelian scalar clockwork model.
Just as in the abelian case, we see that coupling the Goldstone bosons to fields on the $y=0$ brane results in a clockwork suppressed coupling $\sim e^{-kL}$ which we identify with $q^{-N}$ from the discrete models.
Working in the effective theory of the action in \ref{5DNaction}, by integrating out the radial mode, we obtain
\begin{align}
S_{\textsc{na}}^{\rm EFT}&=\int\! d^4x\int_0^L\!dy~e^{2ky}\frac{v^2}{2L}\text{Tr}\left[|\partial_MU|^2\right], \nonumber \\
&=\int\! d^4x\int_0^L\!dy~e^{2ky}\frac{v^2}{2L}\text{Tr}\Big[\frac{1}{2}\partial^{\mu}\pi^{\hat a}\partial_{\mu}\pi^{\hat a}-\frac{1}{2}\partial_5\pi^{\hat a}\partial_5\pi^{\hat a}\Big]
\end{align}
where $U(x,y)\!=\!\exp\left(i\sqrt2T^{\hat a}\pi^{\hat a}/v\right)$.
Discretizing this action results in the a 4D action which at the quadratic level has the same form as that in the discrete non-abelian clockwork set-up described in the previous section.
However here the higher order terms are not present, as in the continuum limit these terms do not survive.

\subsection{Non-abelian clockwork from a 5D gauge theory}

In complete analogy with section \ref{abelianA5} we can implement a clockwork mechanism for a scalar multiplet, charged under some local or global symmetry, from a gauge theory in the bulk of the LD model.
We start with a non-abelian bulk gauge symmetry $\mathcal{G}$, with the following action, up to quadratic terms
\beq
S_{\textsc{gna}}=\int\! d^4x\int_0^L\!dy~e^{2ky}\left[-\frac{1}{4}F^{MN}_\textsc{a}F^{\textsc{a}}_{MN}+\frac{1}{2\xi}\left(\partial_{\mu}A_\textsc{a}^{\mu}+\xi e^{-2ky}(\partial_5e^{2ky}A^\textsc{a}_5)\right)^2\right]
\eeq 
where $\textsc{a}\!=\!1,\ldots,p$, labels the generators of the gauge symmetry.
On the $y\!=\!0$ brane we can break this gauge symmetry, via a choice of orbifold boundary conditions, to a subgroup $\mathcal{H}$.
Then on the $y\!=\!L$ brane we break the gauge symmetry yet again, this time to a subgroup of $\mathcal{H}$ that we call $\tilde{\mathcal{H}}$.
We label the generators in $\tilde{\mathcal{H}}$ by $\tilde a$, the generators in $\mathcal{G}/\mathcal{H}$ by $\hat{a}$, and all others by $a$.
Just as in section \ref{abelianA5} we break this bulk symmetry by imposing Dirichlet boundary conditions on the appropriate brane.
The massless zero mode spectrum then consists of dim($\tilde{\mathcal{H}}$) massless gauge fields transforming in the adjoint of $\tilde{\mathcal{H}}$ and dim($\mathcal{G}$)$-$dim($\mathcal{H}$) massless scalars transforming in the $\mathcal{G}/\mathcal{H}$ coset.
The KK spectrum consists of dim($\mathcal{G}$) massive KK gauge modes transforming in the adjoint of $\mathcal{G}$.
The massless scalars in the $\mathcal{G}/\mathcal{H}$ coset then represent the Goldstone bosons of the spontaneous symmetry breaking of $\mathcal{G}$ to $\mathcal{H}$.
And the massless gauge fields represent an external gauging of a subset of the global symmetry.
This structure is common to other extra dimensional models such as gauge-Higgs unification models \cite{Hosotani:1983xw,Contino:2003ve,Agashe:2003zs,Agashe:2004rs,Hosotani:2005nz}.

Applying the same techniques as in section \ref{abelianA5} we find that the physical 5D profiles of these fields are given by
\begin{align} \label{nonabelianprofiles}
f^{5}_0(y)=&\frac{e^{-ky}}{N_0^5},~~~~N_0^5=\sqrt{\tfrac{2kL}{1-e^{-2kL}}}	\non\\
f^{a}_0(y)=&\frac{e^{ky}}{N_0^a},~~~~N_0^a=\sqrt{\tfrac{2kL}{e^{2kL}-1}}	\non\\
f^a_n(y)=&\frac{\sqrt{2}n\pi}{m^a_nL}\left(\cos\left(\beta^a_n y\right)+\tfrac{k}{\beta^{a}_n}\sin\left(\beta^a_n y\right)\right)	\non\\
f^{\hat{a}}_n(y)=&\frac{\sqrt{2}n\pi}{m^{\hat{a}}_nL}\sin\left(\beta_n^{\hat{a}} y\right)	\non\\
f^{\tilde{a}}_n(y)=&\frac{\sqrt{2}n\pi}{m^{\tilde{a}}_nL}\sin\left(\beta_n^{\tilde{a}} y\right)	
\end{align}
where $f_0^{5}$ is the profile for the massless $A_5$ modes and $f^{a}_0$ for the massless gauge fields.
The mass spectrum of the KK modes is determined by the $\beta_n$ functions, these are fixed by the boundary conditions to be
\begin{align}
\beta_n^{\tilde{a}}=&~\beta_n^{\hat{a}}=\frac{n\pi}{L}	\non\\
\tan\left(\beta_n^{a}L\right)=&-\frac{\beta_n^{a}}{k}.
\end{align}
Therefore we find that the mass spectrum of KK modes in $\tilde{\mathcal{H}}$ or $\mathcal{G}/\mathcal{H}$ have the usual clockwork mass spectra, that is
\beq
m_n^2=k^2+\left(\frac{n\pi}{L}\right)^2.
\eeq
However other fields, which have no $A_{\mu}$ or $A_5$ zero mode, have a slightly perturbed mass spectra.
The mass gap is still $\sim\!k$, however the mass splittings are altered and the effect is larger for smaller $kL$ values, i.e. smaller clockworking.

An interesting feature of the physical localisations in Eq.~\ref{nonabelianprofiles} is that
\beq
f_0^5(y)=f_0^{\tilde a}(y)\Big|_{k\rightarrow-k}.
\eeq
Depending on the sign of $k$, either the $A_5^0$ or the $A_{\mu}^0$ couplings will get suppressed by the clockwork mechanism.
In switching from $k>0$ to $k<0$ the `IR' brane for the $A_5$ mode remains at $y\!=\!0$.  External fields coupled to the bulk fields at this point (with $k<0$) will find that they have suppressed couplings to the $A_5$ zero mode and $\mathcal{O}(1)$ couplings to the zero mode gauge field.
The last feature of this 5D model we will look at is the quartic couplings.
For each of the relevant effective couplings we have
\begin{align}
g_4^2A_5^0A_5^0A^{\mu,n}A_{\mu}^n,& 	&g_4^{\prime}&\sim\frac{kg_5}{(N_0^5)^2},		\non\\
g_4^2A_5^0A_5^0A^{\mu,0}A_{\mu}^0,&	&g_4&\sim \frac{kg_5}{(N_0^a)^2},		\non\\
g^{\prime 2}_4A^{\mu,n}A_{\mu}^nA^{\nu,0}A_{\nu}^0,&	&g^{\prime}_4&\sim \frac{kg_5}{(N_0^a)^2},	\non\\
g^{\prime 2}_4A^{\mu,n}A_{\mu}^nA^{\nu,n}A_{\nu}^n,&	&g^{\prime}_4&\sim \frac{kg_5}{(N_n)^2}.
\end{align}
We see that the couplings between the KK modes and the zero mode are suppressed to either the $A_5$ mode or the $A_{\mu}$ mode, depending on the sign of $k$.
The coupling between the $A_{\mu}$ and $A_5$ modes is the most interesting, as it is only suppressed when $k\!>\!0$, in which case the couplings of the $A_5$ scalar do not get suppressed by the clockwork mechanism.
However when we would like to use the clockwork mechanism to suppress the $A_5$ interactions we are required to choose $k\!<\!0$ and thus the effective gauge couplings are no longer suppressed.
The deconstruction of this model follows in a way which is exactly analogous to the abelian case.
Except that now there are mixed boundary conditions and massless gauge fields will remain for the subgroup of generators which are not broken on the branes.

\section{Conclusions}

In this paper we investigated several aspects of the clockwork mechanism in relation to scalar fields.
In Sec.~\ref{quartics} we demonstrated that the higher order interactions between gears in the discrete clockwork models for scalar fields have a very particular parity-like structure that persists to all orders in the expansion of the potential.  This structure is reminiscent of the KK-parity in interactions of KK modes in a flat extra dimension.  However the large-N limit of the clockwork models is not described by a linear dilaton extra dimension rather than a flat extra dimension, and the higher order interactions involving gears actually vanish in this large-N limit.

In Sec.~\ref{abeliansec} we studied two realizations of the clockwork mechanism in the LD extra dimensional model.  In the first we have a complex scalar field with a quartic potential in the bulk and no brane potentials.  The absence of brane potentials implies that the 5D VEV induced by the bulk potentials has a flat 5D profile.  We then studied the KK decompositions of the radial and longitudinal fluctuations and found two particularly interesting results.  The first being that, again due to the absence of brane potentials and presence of a bulk mass for the radial mode, the KK tower of the radial mode does not contain a zero mode.  Only the KK tower for the longitudinal fluctuation has a zero mode, this is because the field has no non-derivative interactions in the 5D bulk.  This zero mode is identified with the Goldstone boson of the spontaneously broken bulk $U(1)$ global symmetry.  The second interesting feature is the presence of a mass gap between the KK modes of the longitudinal and radial KK modes.  The lightest KK mode of the longitudinal fluctuation is determined only by the curvature of the LD model, whereas the lightest mode in the KK tower of the radial fluctuation is determined both by the curvature and by the bulk mass of the complex scalar field.  Thus a hierarchy in these two parameters creates a separation of scales between the longitudinal and radial KK modes.
We then studied the properties of $A_5$ scalar zero modes arising from a bulk $U(1)$ gauge field in the LD model.  Using boundary conditions we then break this $U(1)$ symmetry on the branes, resulting in a massless component for only the $A_5$ component of the bulk gauge field.  This $A_5$ component is identified with the Goldstone boson of the spontaneously broken $U(1)$ global symmetry.
In this model the gears are obviously spin-1 modes and have the exact same mass spectra as the gears in the bulk scalar model.
And importantly, the couplings of these $A_5$ modes to fields on the $y=0$ brane are shown to also have clockwork suppressed interactions when $k<0$.
This is due to the exponential localization of the physical 5D profiles of these fields which, in fact, has exactly the opposite localization properties as the bulk scalar zero mode or would-be bulk spin-1 zero mode.
This can clearly be seen from the deconstructed model which we present at the end of Sec.~\ref{abeliansec}.

In Sec.~\ref{Non-abelian clockwork} we discuss a non-abelian scalar clockwork model based on an $SU(\mathcal{N})_L\times SU(\mathcal{N})_R$ global symmetry.  A simple generalization of the abelian clockwork model does not work.  The potential for the gears contains higher order terms, beginning at the quartic order, which explicitly break the global symmetry.  However, the higher order terms go to zero in the large-N limit and therefore this continuum limit description is free of the terms which break the global symmetry.  We therefore move on to study two 5D realizations of a non-abelian clockwork mechanism for scalar fields.  
These are exactly analogous to those found in Sec.~\ref{abeliansec} for the abelian model, except that now the components of the bulk gauge field have self interactions, which we discuss in some detail.
The deconstruction of the 5D non-abelian gauge field follows in a way which is completely analogous to the abelian case, resulting in a 4D model in which the massless multiplet of Goldstone bosons is asymmetrically distributed along the lattice.  This asymmetric localization leads to clockwork suppressed interactions for the Goldstone bosons.
These features may prove useful in BSM model-building scenarios.

\paragraph{Note Added:} After this paper appeared online there have since been various applications of the clockwork mechanism, in particular to dark matter \cite{Hambye:2016qkf}, axion models \cite{Coy:2017yex} and linear dilaton inflation \cite{Im:2017eju}.

\section*{Acknowledgements}
We would like to thank Matthew McCullough for an in-depth discussion on the different forms of the clockwork mechanism and for his comments on the draft.  We would also like to thank Pedro Schwaller and Yevgeny Kats for useful comments. 
The work of AA was partially supported by the National Science Centre (Poland) research projects, decision DEC-2014/13/B/ST2/03969 and DEC-2014/15/B/ST2/00108, and BMD was supported by EPSRC Grant EP/P005217/1. AA is also grateful to the Mainz Institute for Theoretical Physics (MITP) for hospitality and support during the completion of this work. 

\appendix

\section{Clockwork interactions from higher derivative terms in 5D} 
\label{appendixA}

In this appendix we will look at higher order derivative interactions in the 5D abelian clockwork model. We will work purely in the EFT framework where the UV completion of such scenario is not required for the sake of our argument. 
Including up to four derivative terms in the non-linear sigma expansion $U(x,y)\equiv \exp(i\pi(x,y)/f_{\pi})$ defined in \eqref{phi_decomposition}, we have
\begin{align} \label{5Dnls}
\delta S=\int \!d^4x\int_0^L\!dy\,\frac{w(y)}{L}\Big[ &c_1\big(\partial_MU^{\dagger}\partial^MU\big)^2+c_2\big(\partial_MU^\dag\partial_NU^\dag\partial^MU\partial^NU\big)\Big] + \ldots\,,
\end{align}
where $c_{1,2}$ are some dimensionless constants, $f_\pi=\sqrt{L}\langle\phi\rangle$ is the VEV of the $\phi$ field ($[f_\pi]=1$) and $w(y)$ is a $y$-dependent localisation factor. 
This factor $w(y)$ has been kept general as it may be possible that these terms could arise with a non-trivial dependence along the extra dimension.

After expanding in the KK basis \eqref{piKKmodes} and integrating over the extra dimension to obtain an effective theory, where the $\partial_5U$ terms result in self-interactions among the $\pi_n$ fields, whereas the $\partial_{\mu}U$ terms give the usual 4D derivative expansion.
We can rewrite the above action in terms of $\pi$ KK-modes \eqref{piKKmodes} as
 \begin{align}
 \delta S&\supset \int\! d^4x\Big[ \frac{\tilde{c}^{klmn}_{1}}{f_\pi^4}\big(\partial_\mu \pi_k\,\partial^\mu \pi_l\,\partial_\nu \pi_m\,\partial^\nu \pi_n\big) +\frac{\tilde{c}^{klmn}_{2}}{f_\pi^4}\big(\partial_\mu \pi_k \,\partial_\nu \pi_l\, \partial^\mu \pi_m\,\partial^\nu \pi_n\big) + \ldots~\Big],
 \end{align}
where the coefficients of the $\partial_\mu\pi_n$ terms in the above action are of the form
\beq
\tilde{c}^{klmn}_{1,2}=\frac{4c_{1,2}}{L}\int_0^L\!dy\,w(y)e^{-4ky}f^\pi_k(y)f^\pi_l(y)f^\pi_m(y)f^\pi_n(y)\,,
\eeq
and $\tilde{c}^{klmn}_{1,2}$ are dimensionless. 
For the zero-mode four-derivative interaction we have
\beq
\tilde{c}^{0000}_{1,2}=\frac{4c_{1,2}}{L}\int_0^L\!dy~e^{(r-4)ky}\big(f^\pi_0(y)\big)^4\simeq c_{1,2}\frac{16kL}{r}e^{(r-4)kL}
\eeq
where we have chosen $w(y)\!=\!e^{rky}$ and $f^\pi_0(y)\!\simeq\!\sqrt{2kL}\,e^{k(y-L)}$.
For higher order terms involving the KK modes we do not get the same $e^{(r-4)kL}$ factor in the effective $\tilde{c}_{1,2}$ couplings.  We find that as long as $r\leq4$ the couplings are $\mathcal{O}(1)$.

The 5D action in Eq. \ref{5Dnls} also gives rise to non-derivative interactions for the KK modes in the effective theory,
\begin{align}
\delta S\supset \int \!d^4x\int_0^L\!dy~\frac{w(y)}{L}\Big[ c_3\Big(\left(\partial_5U\right)^{\dagger}\left(\partial_5U\right)\Big)^2 \Big] + \ldots\,,
\end{align}
where $c_3\equiv  (c_1+c_2)$ and the ellipsis represent all the other interactions in which the $\partial_\mu$ and $\partial_5$ terms are mixed but we only consider the above as an example.
Expanding the above interactions in terms of the KK modes of $\pi(x,y)$ Eq.~\eqref{piKKmodes}, we have
\begin{align}
\delta S\!&\supset\! \int \!\!d^4x\int_0^L\!\!dy\frac{w(y)}{L}\frac{c_3}{f_\pi^4}\left[\!\big(\partial_5e^{-ky}f^\pi_k\big)\!\big(\partial_5e^{-ky}f^\pi_l\big)\!\big(\partial_5e^{-ky}f^\pi_m\big)\!\big(\partial_5e^{-ky}f^\pi_n\big)\!\right]\pi_k\pi_l\pi_m\pi_n \,, \notag\\
&=\int \!\!d^4x~ \tilde c_3^{klmn}\,\pi_k\,\pi_l\,\pi_m\,\pi_n \,,
\end{align}
where $\tilde c_3^{klmn}$ is a dimensionless effective coupling constant. 
One can see already that these terms will not involve the zero-mode, which is consistent with the fact that we interpret the zero-mode as the Goldstone boson of the broken global $U(1)$ symmetry.
The 5D profiles of the gears take a simpler form under the derivative
\beq
\partial_5 \big[e^{-ky}f^\pi_k(y)\big]=-\sqrt2 m_ke^{-ky}\sin\big(\tfrac{k\pi y}{L}\big)
\eeq
thus the effective coupling constant can be written as
\begin{align}
\tilde{c}_3^{klmn}=4\frac{c_3}{f_\pi^4}m_km_lm_m m_n\int_0^Ldy~\frac{w(y)}{L}e^{-4ky}\sin\big(\tfrac{k\pi y}{L}\big) \sin\big(\tfrac{l\pi y}{L}\big)\sin\big(\tfrac{m\pi y}{L}\big)\sin\big(\tfrac{n\pi y}{L}\big).
\end{align}
If we choose $w(y)=e^{4ky}$ we are left with an interaction which has the exact same structure as that found in the discrete case, i.e. Eqs.~\eqref{Cklmn} and \eqref{Cklmn2}.
In this case the KK modes have a discrete KK parity symmetry similar to that found in flat extra dimensional models.
This happens because the $w(y)=e^{4ky}$ cancels off the warping factors from the profiles of the scalar fields.

\section{Bulk gauge fields in the linear dilaton geometry}
\label{gauge_LDM}
In this Appendix we work all the properties of the gauge fields in the LD background geometry in the Jordan and Einstein frames. 
\subsection{Jordan frame}
We start with the gauge fields in the {\it Jordan frame} of LD geometry and the action can be written as
\begin{align}
S_A &= \int \! d^4x\int_0^Ldy e^{2ky} \left\{  -\frac{1}{4}
  g^{MN} g^{KL} F_{MK} F_{NL}  \right\} + S_{\rm GF}~,
\label{GaugeAction}
\end{align}
 where all the contractions are w.r.t. the Minkowski metric and we will write the gauge-fixing term $S_{\rm GF}$ later below.
In order to find the bulk equation of motion for the KK-modes we will be interested in the quadratic part of the
action. Hence, there is no distinction between the abelian and
non-abelian cases, and we can safely consider $F_{MN} = \partial_M A_N -
\partial_N A_M$, where $M,N,\ldots\!=\!0,1,2,3,5$.  In order to find the gauge fixing term $S_{\rm GF}$ let us expand the gauge kinetic term:
\begin{align}
 &\int \! d^4x\int_0^L\!dy e^{2ky} \left\{
-\frac{1}{4} F_{\mu\nu} F^{\mu\nu} + \frac{1}{2} F_{\mu5} F^{\mu}_5
\right\}
\nonumber \\ 
&=
\int \! d^4x\int_0^L\!dy e^{2ky} \bigg\{\!\!-\frac{1}{4} F_{\mu\nu} F^{\mu\nu} + \frac{1}{2}\partial_{5} A_\mu \partial_{5} A^\mu + \frac{1}{2} 
\partial_{\mu} A_5 \partial^{\mu} A_5 
- \partial_{5}A_\mu\partial_{\mu} A_5
\bigg\},\notag\\
&=
\int \! d^4x\int_0^L\!dy e^{2ky} \bigg\{\!\!-\frac{1}{4} F_{\mu\nu} F^{\mu\nu} + \frac{1}{2}\partial_{5} A_\mu \partial_{5} A^\mu + \frac{1}{2} 
\partial_{\mu} A_5 \partial^{\mu} A_5 
- e^{-2ky} \partial_{5} \! \left[ e^{2ky} A_5 \right] \partial_{\mu} A^\mu
\bigg\} \notag\\
&~+\int\!d^4x ~e^{2ky} A_5 \partial_{\mu} A^\mu\Big\vert_{y\!=\!0}^{y\!=\!L},
\end{align}
where in the last step we integrated by parts the last term of second line w.r.t.  $x^\mu$ and $y$. The
boundary terms corresponding to the $x^\mu$ integration are but to zero as usual, however we keep the boundary terms for the $y$ integration. Note that there is a bulk kinetic mixing of the $A_\mu$ and $A_5$
components, which also results in mixing in the EOM. Hence the gauge-fixing
action should be chosen such that to cancel this term, i.e.
\begin{align}
S_{\rm GF} &=-\int \! d^4x\int_0^L\!dy e^{2ky}\frac{1}{2\xi} \Big(
\partial_\mu A^\mu - \xi \, e^{-2ky} \partial_5 \left[ e^{2ky} A_5 \right]
\Big)^2,
\label{SS}
\end{align}
where $\xi$ is a gauge-fixing parameter. 
After finding the gauge fixing part of the action, we can write down the total gauge action in the LD geometry as
\begin{align}
S &=\int \! d^4x\int_0^L\!dy e^{2ky}\,\bigg\{
-\frac{1}{4} F_{\mu\nu} F^{\mu\nu} -\frac{1}{2\xi} \left(
\partial_\mu A^\mu \right)^2 +
\frac{1}{2} \partial_{5} A_\mu \partial_{5} A^\mu
\nonumber \\ 
&  \hsp+ \frac{1}{2}
\partial_{\mu} A_5 \partial^{\mu} A_5
- \frac{1}{2} \xi \left( e^{-2ky}\partial_5 \left[ e^{2ky} A_5 \right]
\right)^2
\bigg\}+\int\!d^4x ~e^{2ky} A_5 \partial_{\mu} A^\mu\Big\vert_{y\!=\!0}^{y\!=\!L}. \label{gaugeActionFull}
\end{align}

In order to get the bulk equations of motion for the gauge field $A_\mu$ and the gauge-scalar $A_5$, let us now consider the variation of the above action w.r.t. to $A_\mu$ and $A_5$,
\begin{align}
\delta S^{\rm bulk}_{A_\mu} &=\!\int \! d^4x\int_0^L\!\!dy e^{2ky}\delta A_\mu \Big\{
\Big[ \eta^{\mu\nu} \Box - \Big(\! 1 -\! \frac{1}{\xi} \!\Big)
\partial^\mu \partial^\nu\! \Big]\!
-\eta^{\mu\nu} e^{-2ky}\partial_5 \, e^{2ky} \partial_5
\Big\} A_\nu,		\label{dAmuB}\\ 
\delta S^{\rm boundary}_{A_\mu} &= \int \! d^4x \, e^{2ky} \delta A_\mu \Big\{\partial_5 A^\mu - \partial^\mu \! A_5
\Big\} \Big\vert_{y=0}^{y=L}~,		\label{dAmuS}\\
\delta S^{\rm bulk}_{A_5} &= \int \! d^4x\int_0^L\!dy e^{2ky}\, \delta A_5 \Big\{
-\Box + \xi \, \partial_5 e^{-2ky}\partial_5 \, e^{2ky} \Big\} A_5~, 	\label{dA5B}\\ 
\delta S^{\rm boundary}_{A_5} &= \int \! d^4x e^{2ky}\, \delta A_5 \Big\{\partial_\mu A^\mu-
\xi \, \partial_5 \big[ e^{2ky} A_5 \big]\Big\} 
\Big\vert_{y=0}^{y=L}~.	\label{dA5S}
\end{align}
where $\Box \equiv \eta^{\mu\nu} \partial_\mu \partial_\nu$, and we performed an integrations by parts on $x^\mu$ for the boundary term of Eq.~\eqref{gaugeActionFull} in Eq.~\eqref{dAmuS}. Now its straightforward to write the bulk EOM for the $A_\mu$ and $A_5$ fields:
\begin{align}
 \Big[ \eta^{\mu\nu} \Box - \Big( 1 - \frac{1}{\xi} \Big)
\partial^\mu \partial^\nu \Big] A_\nu
- \eta^{\mu\nu}e^{-2ky}\partial_5 \! \Big[ e^{2ky} \partial_5 A_\nu \Big]&=0~,
\\ 
 \Box A_5 - \xi \, \partial_5\big[e^{-2ky}\partial_5 \! \big( e^{2ky} A_5 \big)\big]&=0~.
\end{align}
Along with above EOMs for the $A_\mu$ and $A_5$ fields, we can have the following sets of boundary conditions,
\begin{align}
\partial_5 A_\mu\big\vert&=0, \lsp\text{which implies} \!\!\!\Lsp A_5\big\vert=0 ,		\\
A_\mu\big\vert&=0, \lsp\text{which implies} \hsp \partial_5(e^{2k y}A_5)\big\vert=0,	
\end{align}
where we note that the boundary conditions for the $A_\mu$ and $A_5$ are not independent of each other, for example choose Neumann b.c. for $A_\mu$ at $y=0,L$, leads to zero-mode for the $A_\mu$ field, whereas there will be no $A_5$ zero-mode. However, in this section we are interested in the zero-mode of the $A_5$ mode, therefore we choose Dirichlet b.c. for $A_\mu$ at $y=0,L$, which implies Neumann b.c. for the $A_5$ mode. Hence there would be a zero-mode for the $A_5$ scalar. 
\subsubsection*{KK-decomposition of gauge fields}

We can write the 5D gauge fields $A_{\mu,5}(x,y)$ as the Fourier-mode decomposition or commonly known as KK decomposition, as
\beq
A_{\mu(5)}(x^\mu,y)= \frac{e^{-ky}}{\sqrt{L}} \sum_{n=0}^\infty A_{\mu(5),n}(x^\mu) f^{A(5)}_n(y)~,
\label{gaugeKKA}
\eeq
where $A_{\mu(5),n}(x^\mu)$ are the 4D KK-modes and $f^{A,5}_n(y)$ are the 5D KK wave-functions. Note that in the above decomposition a factor $1/{\sqrt{L}}$ take care of the dimensionality, such that $[A_n]\!=\!1$ and $[f_n]\!=\!0$. Moreover, we have an explicit factor of
$e^{-ky}$, which makes the kinetic terms for $A_{\mu(5),n}$ canonically normalized and in this decomposition KK wave-functions determines the localization properties
of the KK-modes in the proper distance. With the above decomposition, it is straight forward to see from the requirement to have the kinetic terms of KK-modes canonically normalized, the KK wave-functions satisfy,
\begin{align}
\frac{1}{L} \int_0^L \! dy \, f^{A(5)}_m f^{A(5)}_n &=\delta_{mn}~.
\label{gaugeKKNorm}
\end{align}

The form of the KK wave-functions is specified by the bulk EOM for the 5D $A_{\mu,5}$ fields (we will work in the Feynman gauge $\xi\!=\!1$), i.e.
\begin{align}
e^{-ky}\partial_5 \left[ e^{2 ky} \partial_5 (e^{-ky}f^A_n )\right] + m^2_{A,n} f^A_n &=
0~,
\label{eomfA}\\
e^{ky}\partial_5 \left[ e^{-2 ky}\partial_5 ( e^{ky} f^5_n)\right]  + m^2_{A,n} f^5_n &=0~.
\label{eomf5}
\end{align}
Note that the above two equations for the $f^A_n$ and $f^5_n$ are not completely independent since KK-mass $m^2_{A,n}$ is same for both cases, hence finding the solution for $f^A_n$ with $m_{A,n}^2\neq0$ would also give a solution for the $f^5_n$ as
\beq
f^5_n (y)= \frac{ e^{ky}}{m_{A,n}}\partial_5\big[e^{-ky} f^A_n(y)\big]~.	\label{A5profile}
\eeq
This is due to the manifestation of Higgs mechanism at each KK-mode level, i.e. KK gauge fields $A_{\mu,n}$ are acquiring mass $m_n$ by `eating' the $A_{5,n}$ scalar mode. 

Let us start with the zero-mode wave-functions for the $A_{\mu,5}$.
For the Neumann b.c.'s for $A_\mu$, i.e. 
\beq
\partial_5 e^{-ky} f^A_n \Big\vert_{0,L}=0, \Lsp 
f^5_n \Big\vert_{0,L} = 0.
\eeq
The solutions of for zero-modes would be, 
\beq
f^A_0(y)=N^A_0e^{k(y-L)}, \Lsp 
f^5_0(y) = 0,
\eeq
where $N^A_0$ is fixed by the normalization condition Eq.~\eqref{gaugeKKNorm}, 
\beq
N^A_0=\sqrt{\tfrac{2kL}{1-e^{-2kL}}}.
\eeq
The solutions for the non-zero KK wave-functions are, 
\beq
f^A_n(y)=\tfrac{\sqrt{2}n\pi}{m_{A,n}L}\left[\cos\left(\tfrac{n\pi y}L\right)+\tfrac{kL}{n\pi}\sin\left(\tfrac{n\pi y}{L}\right)\right], \lsp f^5_n(y)=-\sqrt2 \sin\left(\tfrac{n\pi y}{L}\right).
\eeq
Now let us consider the Dirichlet b.c.'s for $A_\mu$ (Neumann b.c.'s for $A_5$, ), i.e. 
\beq
f^A_n \Big\vert_{0,L}=0, \Lsp 
\partial_5 e^{ky} f^5_n \Big\vert_{0,L} = 0.
\eeq
These boundary conditions leads to solutions of for zero-modes as 
\beq
f^A_0(y)=0, \Lsp 
f^5_0(y) = N^5_0e^{-ky},
\eeq
where $N^5_0$ is fixed by the normalization condition Eq.~\eqref{gaugeKKNorm}, 
\beq
N^5_0=\sqrt{\tfrac{2kL}{1-e^{2kL}}}.
\eeq
The solutions for non-zero KK profiles for these boundary conditions are, 
\beq
f^A_n(y)=\sqrt2 \sin\left(\tfrac{n\pi y}{L}\right), \lsp f^5_n(y)=\tfrac{\sqrt{2}n\pi}{m_{A,n}L}\cos\left(\tfrac{n\pi y}L\right)\,.
\eeq


\subsection{Einstein frame}
For the case of bulk gauge fields, in the Einstein frame, we will consider the following metric which corresponds to the canonically normalized Einstein-Hilbert 5D action for gravity and as well canonically normalized 5D matter (gauge field) action,
\beq
g^{(\rm E)}_{MN}=e^{2 A(y)}\eta_{MN},	\lsp\text{where for LD geometry}\hsp A(y)=\frac23ky. \label{metricE}
\eeq
In the Einstein frame the action for bulk gauge fields can be written as:
\begin{align}
S^{(\rm E)}_{A} &= \int \! d^4x\int_0^L\!dy \sqrt{g^{(\rm E)}} \left\{  -\frac{1}{4}
  g_{(\rm E)}^{MN} g_{(\rm E)}^{KL} F_{MK} F_{NL}  \right\} + S^{(\rm E)}_{\rm GF}~,\notag\\
&= \int \! d^4x\int_0^L\!dy \, e^{5A(y)} \left\{
-\frac{1}{4} \, e^{-4A(y)}  F_{\mu\nu} F^{\mu\nu} + \frac{1}{2} \,
e^{4A(y)} F_{\mu5} F^{\mu}_5
\right\}+ S^{(\rm E)}_{\rm GF}
\nonumber \\
&=
\int \! d^4x\int_0^L\!dy\, e^{A(y)}\bigg\{
-\frac{1}{4} F_{\mu\nu} F^{\mu\nu}
+ \frac{1}{2} \partial_{5} A_\mu \partial_{5} A^\mu + \frac{1}{2}
\partial_{\mu} A_5 \partial^{\mu} A_5	\notag\\
&\Lsp\hsp- e^{-A(y)}\partial_{5} \! \left[ e^{A(y)} A_5 \right] \partial_{\mu} A^\mu
\bigg\}+\int \! d^4x \left. e^{A(y)} A_5 \partial_{\mu} A^\mu
\right|_{y=0}^{y=L}+ S^{(\rm E)}_{\rm GF}~,
\label{gaugeActionE}
\end{align}
where in the last two steps all the contractions are now done with the Minkowski metric.
Above the gauge-fixing action $S^{(\rm E)}_{\rm GF}$ is,
\begin{align}
S^{(\rm E)}_{\rm GF} &= -\frac{1}{2\xi} \int \! d^4x\int_0^L\!dy e^{A(y)}\Big(
\partial_\mu A^\mu - \xi \, e^{-A(y)}\,\partial_5 \left[ e^{A(y)} A_5 \right]
\Big)^2~,
\label{SE_GF}
\end{align}
With the above gauge fixing action, we get the following for of the total action,
\begin{align}
S^{\rm (E)}_A &=\int \! d^4x\int_0^L\!dy e^{A(y)}\,\bigg\{
-\frac{1}{4} F_{\mu\nu} F^{\mu\nu} -\frac{1}{2\xi} \left(
\partial_\mu A^\mu \right)^2 +
\frac{1}{2} \partial_{5} A_\mu \partial_{5} A^\mu
\nonumber \\ 
&  \hsp+ \frac{1}{2}
\partial_{\mu} A_5 \partial^{\mu} A_5
- \frac{1}{2} \xi \left( e^{-A(y)}\partial_5 \left[ e^{A(y)} A_5 \right]
\right)^2
\bigg\}+\int\!d^4x ~e^{A(y)} A_5 \partial_{\mu} A^\mu\Big\vert_{y\!=\!0}^{y\!=\!L}. \label{gaugeActionFullE}
\end{align}
Note that the above action is exactly as in the Jordan frame, except the exponential factors ($2ky\to A(y)=2ky/3$). Hence all the conclusion would remain unchanged up to $k\to k/3$.


\providecommand{\href}[2]{#2}\begingroup\raggedright\endgroup

\end{document}